\documentstyle[12pt,epsf]{article}
\newcommand{\be}{\begin{equation}}
\newcommand{\bea}{\begin{eqnarray}}
\newcommand{\eea}{\end{eqnarray}}
\newcommand{\beas}{\begin{eqnarray*}}
\newcommand{\eeas}{\end{eqnarray*}}
\newcommand{\ba}{\begin{array}}
\newcommand{\ea}{\end{array}}
\newcommand{\ee}{\end{equation}}

\newcommand{\tr}{{\rm Tr}\ }
\newcommand{\str}{{\rm STr}\ }
\def\identity{{\rlap{1} \hskip 1.6pt \hbox{1}}}
\newcommand{\nbox}{{\,\lower0.9pt\vbox{\hrule \hbox{\vrule height 0.2 cm \hskip
0.2 cm \vrule height 0.2 cm}\hrule}\,}}

\renewcommand{\a}{\alpha}
\renewcommand{\b}{\beta}

\newcommand{\pa}{\partial}
 
\newcommand{\e}{\epsilon}

\newcommand{\r}{\rho}

\newcommand\text[1]{\rm #1}

\newcommand\cf{{\cal F}}
\newcommand\co{{\cal O}}
\newcommand\ca{{\cal A}}
\newcommand\cR{{\cal R}}
\newcommand\cp{{\cal P}}

\def\da{{\dot{\alpha}}}
\def\db{{\dot{\beta}}}

\def\bX{{\bar{X}}}
\def\bc{{\bar{\chi}}}
\def\bl{{\bar{\lambda}}}
\def\bt{{\bar{\theta}}}
\def\P{{\Phi}}
\def\bP{{\bar{\Phi}}}

\newcommand{\CC}{\hbox{\xiiss C\kern-.4emI}}
\newcommand{\RR}{\hbox{\xiiss R\kern-.45emI}}
\newcommand{\ZZ}{\hbox{\xiiss Z\kern-.4emZ}}
\newcommand{\CCs}{\hbox{\ixss C\kern-.4emI}}
\newcommand{\ZZs}{\hbox{\ixss Z\kern-.4emZ}}
\newcommand{\pasl}{\pa\kern-.55em /}
\def\href#1#2{#2}

\def\a{\alpha}
\begin{document}
\begin{titlepage}
\hfill
\vbox{
    \halign{#\hfil         \cr
           hep-th/0305145 \cr
           SU-ITP-03/07 \cr
           } % end of \halign
      }  % end of \vbox
\vspace*{20mm}
\begin{center}
{\Large Blending Local Symmetries With Matrix Nonlocality \\
In D-brane Effective Actions}

\vspace*{15mm}
\vspace*{1mm}
%{Authors}
{Mark Van Raamsdonk} 

\vspace*{1cm}

{Department of Physics, Stanford University\\
382 via Pueblo Mall, Stanford CA 94305-4060, USA}

\vspace*{0.4cm}

{Department of Physics and Astronomy,
University of British Columbia\\
6224 Agricultural Road, 
Vancouver, B.C., V6T 1W9, Canada}

\vspace*{1cm}
%%\maketitle
\end{center}

\begin{abstract}

In systems of intersecting branes, we consider sets of directions in which one type of brane is pointlike, with transverse fluctuations described by matrix coordinates $X$, and the other set of branes is space-filling, with a local symmetry associated to its worldvolume gauge field. Under this symmetry, massless fields associated with p-p' strings should transform in the fundamental representation, $\Phi \to U(X) \Phi$, but this transformation rule is ill-defined when $X$ is a general matrix. In this paper, we make sense of this transformation rule for $\Phi$ and show that imposing gauge invariance using the resulting rule places strong constraints on the effective actions, determining infinite series of terms in the $\alpha'$ expansion. We describe the most general invariant effective actions and note that these are written most simply in terms of covariant objects built from $\Phi$ which transform like fundamental fields living on the whole space-filling brane. Our description leads us to introduce several interesting structures, including Wilson lines from ordinary points to matrix locations, ``pull-backs'' of fields from matrix geometries to ordinary space, and delta functions which localize to matrix configurations. 

\end{abstract}

\end{titlepage}

\vskip 1cm
%\newpage
\section{Introduction}

Local symmetries play a central role in the effective field theories
that describe the physics of currently observable energy scales: much
of the structure of the Standard Model and General Relativity is
determined by gauge symmetries and general coordinate invariance
respectively. One of the most compelling features of string theory is
that its low energy description incorporates precisely these
symmetries, taking the familiar form of gauge theories coupled to
gravity.

In addition to these standard elements, string theory presents some
less familiar characters in its low-energy physics: stable extended
objects, and specifically D-branes. Even at the classical level,
configurations of these objects are described in a novel way; the
transverse positions of N branes are characterized by a set of $N
\times N$ Hermitian matrices \cite{witten}. When these matrices do not commute, the
D-branes of a given dimension take on characteristics of higher
dimensional objects and are described by noncommutative geometries
rather than as ordinary submanifolds of spacetime.

While these noncommutative geometries are characterized by some degree
of ``fuzziness'' or nonlocality, the D-brane degrees of freedom still
interact with the ordinary fields of the theory and specifically the
massless fields associated with local symmetries. Thus, a natural
description of the low-energy effective action for multiple D-branes
in string theory should incorporate both local symmetries and
non-local matrix coordinates in a harmonious way. So far, no such
description has emerged. For example, while the Born-Infeld type
actions describing the coupling of single branes to gravity are
expressed in geometrical language for which coordinate invariance is
manifest, the existing generalizations to multiple D-branes
\cite{myers, tvp} are not generally covariant and must be supplemented
with additional terms (yet to be determined) to realize the symmetry. Optimistically, one might hope that understanding how to realize the local symmetries in D-brane effective actions should determine much of the structure of these additional terms and lead to interesting new physics. 

In one attempt to address these questions, de Boer and Schalm \cite{ds} investigated whether it is possible to add higher-order terms to the leading order actions coupling
D0-branes to a metric such that the result is invariant under general
coordinate transformations with a suitable transformation law for the
D0-brane matrices. Working order-by-order in the D0-brane matrices (up
to sixth order) they actually found many possible actions invariant to the
order that they checked.  Unfortunately, their order-by-order results
for the transformation rules of matrix coordinates and the leading
terms in the invariant action look rather complicated and unenlightening, and
it is not clear how to explicitly generalize them to an all-order invariant
expression such as the Born-Infeld action for single branes.

In this paper, we study a related simpler problem which lies entirely within
the context of open string physics. Consider any intersecting brane
system, and focus on a set of directions for which one set of branes
are space-filling and another set are pointlike. Then there will be a
local gauge symmetry associated with the gauge field on the
space-filling branes. The lowest modes of open strings connecting
the pointlike with the space-filling branes give rise to fields living
on the pointlike branes which should transform in the fundamental
representation of this local gauge symmetry. For a single pointlike
brane, the transformation rule is simply
\be
\label{simp}
\Phi \to U(X) \Phi
\ee
where $U(y)$ is the gauge transformation function on the space-filling
brane and $X$ is the location of pointlike brane. On the other hand,
with more than one pointlike brane, $X$ becomes a matrix, and we
immediately run into a problem of how to interpret (\ref{simp}). 

The main point of this paper is to give a precise prescription for this transformation law and to describe in general the effective actions that are invariant under it. Along the way, we will find that an important role is played by a number of interesting new structures which generalize familiar objects to the realm of noncommutative geometry. A summary and outline of the paper follows. 

In section 2, we take some preliminary steps toward understanding (\ref{simp}) in the case of matrix $X$. We find that order-by-order in $X$, it is possible to define a consistent transformation rule for $\Phi$, which necessarily involves not only $\Phi$, $U(y)$, and $X$, but also the gauge field $A(y)$ on the space-filling brane. Given this transformation rule, we find that simple expressions bilinear in $\Phi$ are not invariant, but by adding additional terms, again order-by-order in $X$, it is possible to come up with invariant expressions. So far, both the transformation rule for $\Phi$ and the gauge invariant expressions that result are messy and unilluminating expansions in $X$. However, we note that the complicated gauge invariant expressions may be written in a simple compact form in terms  of a new object $\cf_a$ (built from $\Phi$, $X$, and $A$) which transforms covariantly as $\cf_a \to U(a) \cf_a$. Here, $a$ is an arbitrary point on the space-filling brane.

In section 3, we describe a more systematic approach the problem. We point out that the action and transformation rules for our problem may be derived directly from those of the simpler situation in which we restrict to constant gauge fields. From this simpler setup, we may return to our original problem in three steps. First, we use a quotient construction analogous to \cite{dm,wati} to compactify on a torus in the direction of the space-filling branes, thereby including winding modes associated to the pointlike branes. Second, we T-dualize on all directions yielding pointlike branes (without winding modes) in the presence of space-filling branes on the torus with a spatially dependent gauge field. Finally, we decompactify the torus. If desired, we may use the quotienting procedure to compactify again (this time with a spatially dependent gauge field) to obtain the full action for pointlike and space-filling branes on the torus with all winding/momentum modes included. Thus, our original problem (and the more complicated problem with pointlike and space-filling branes on a torus) reduces to finding the action and transformation rules in the zero-mode sector and understanding how T-duality acts on the fields of the problem.

In the remainder of the paper, we implement this prescription. In section 3.1, we argue that T-duality should map $\Phi$ into an object with the transformation rules proposed for $\cf_0$ above and henceforth define $\cf = \cf_0$ to be the image of $\Phi$ under T-duality. In section 4, we argue that this definition together with the assumed transformation properties of $\cf$ determine $\cf$ in terms of $\Phi$ up to a certain class of field redefinitions. The relationship (for constant gauge field) is given by a linear operator $\co$,
\[
\cf = \co(A,X) \Phi \equiv \Phi + i A_i \Phi X^i + \dots 
\]
which reduces to 
\[
\co = e^{i A \cdot X} 
\]
when either $X$ or $A$ is abelian. Though we are not able to write a closed form expression for $\co$ in the general case, we provide simple and suggestive equations that determine it and show that to the first several orders in $X$, these equations have a solution unique up to a certain class of field redefinitions. Given this relationship between $\cf$ and $\Phi$, the transformation rules under all gauge and translation symmetries are determined, and we are able to describe a minimal basis of actions for the zero-modes invariant under all symmetries and consistent with T-duality. We find that symmetries determine an infinite series of couplings to the gauge field associated with an arbitrary term in the action for $A=0$. 

In section 5, starting with these invariant zero-mode actions, we follow the quotienting and T-duality procedures to determine the form of invariant actions in the more general situations. For the original situation including pointlike and noncompact space-filling branes, we find that invariant actions may be written in two qualitatively different ways, involving two different covariant objects $\cf_y$ and $\Phi(y)$, both with the same transformation properties. We note that $\cf_y$ (obtained previously in section 2) is something like a parallel transported version of $\Phi$, reducing to $\Phi$ times a straight Wilson line from $X$ to $y$ in the case of Abelian $X$. On the other hand, $\Phi(y)$ is an object that reduces to $\delta(X-y) \Phi$ for Abelian X, or when $A=0$ (with a suitable definition of the delta function for matrix $X$). This is something like a ``projection'' or ``pull-back'' of $\Phi$ from the noncommutative space defined by $X$ to the ordinary worldvolume of the space-filling branes. Using these covariant structures, the action takes the form of an integral over the whole space-filling brane with a matrix delta function (defined in section 4.2) localizing the integral to the vicinity of the pointlike branes. 

For pointlike and space-filling branes on a torus (e.g. in the D1-D5 black hole system), we find that the invariant actions are naturally written as integrals over both the torus and dual torus and involve a new covariant object $\Phi(y, \tilde{y})$ built from $\Phi$. This transforms as
\[
\Phi(y, \tilde{y}) \to U(y) \Phi(y, \tilde{y}) V^{-1}(\tilde{y}) 
\]
under simultaneous gauge transformation on the torus and dual torus and has periodicity properties
\[
\Phi(y + 2 \pi n R, \tilde{y} + 2 \pi m \tilde{R}) = e^{i m y / R} \Phi(y, \tilde{y}) \; .
\]

In section 6, we briefly discuss more general intersecting brane
systems obtained from T-dualizing the systems considered so far, and
as an example, discuss leading corrections to the low-energy Dp-D(p+4)
effective actions determined uniquely by the symmetries and
T-duality. Finally, in section 7, offer a few concluding remarks
including comments on the related problem of coupling D-branes to
gravity in a covariant way.  
 
\section{Preliminaries}

We begin by stating the basic problem in its simplest context. We consider two sets of branes which intersect along some directions. We will only be concerned with the directions occupied by the first set of branes and not the second, and we describe these directions by coordinates $y^i$.  Thus, with respect to these directions, the first set of branes are space-filling while the second set of branes are pointlike, and we will refer to them as such.

On the space-filling branes, we ignore all worldvolume fields apart from the components of the gauge field in the $y^i$ directions, which transform under the local gauge symmetry in the usual way,
\be
\label{gauge}
A_i(y) \to U(y) A_i(y) U^{-1}(y) + i \partial_i U(y) U^{-1}(y) 
\ee
In addition to this gauge field, we consider the matrix coordinates $X^i$ which describe the configuration of the pointlike branes in the $y^i$ directions. These transform under the symmetry group associated with the pointlike branes as
\be
\label{D0sym}
X^i \to V X^i V^{-1}
\ee
Finally, we consider a bifundamental $\Phi$ which arises from the lowest energy modes of strings stretched between the pointlike and space-filling branes.\footnote{In specific examples, there may be a number of bifundamental fields of various types, but for our purposes, we will consider any one of these and call it $\Phi$, suppressing any global symmetry indices it may carry.} Note that $X^i$ and $\Phi$ may be considered to ``live'' on the pointlike branes and do not depend on the coordinates $y^i$. 

Under the pointlike brane symmetry (\ref{D0sym}), $\Phi$ transforms as an antifundamental field 
\[
\Phi \to \Phi V^{-1} \; ,
\]
while under the gauge symmetry of the space-filling brane (\ref{gauge}), $\Phi$ should transform as a fundamental,
\[
\Phi \to U \Phi \; .
\]
However, $U$ here is a function of the coordinates on the
space-filling brane, so this expression is ambiguous unless we specify
where $U$ should be evaluated. For the case of a single pointlike
brane, the answer is clear: since $\Phi$ lives on the pointlike brane,
we should evaluate $U$ at the position of the brane when making a
gauge transformation on $\Phi$,\footnote{This ensures that $\Phi$ is
invariant under gauge transformations $U(y)$ which are trivial in the
neighborhood of the pointlike brane. We are working at a point where
the $\Phi$ fields have no vev and in an $\alpha' \to 0$ limit so that
single branes should really have delta function support.} 
\be
\label{basic}
\Phi \to U(X) \Phi
\ee
On the other hand, for multiple pointlike branes, $X$ becomes a matrix, and it is not clear how we should generalize (\ref{basic}). The problem is that for general matrix configurations, the pointlike branes do not have well-defined positions. The most naive generalization of (\ref{basic}) would be simply to treat $U(X)$ as a Taylor expansion about $X=0$ and define a transformation law
\bea
\label{firsttry}
\Phi \to U \circ \Phi &=& \sum {1 \over n!} \partial_{i_1} \cdots \partial_{i_n} U(0) \Phi X^{i_1} \cdots X^{i_n} \cr && = U(0) \Phi + \partial_i U(0) \Phi X^i + {1 \over 2} \partial_i \partial_j U(0) \Phi X^i X^j + \dots 
\eea
This has no ordering ambiguities since the indices on the derivatives are completely symmetrized. In addition, it gives the correct answer when the matrices $X$ are all diagonal: in this case, the component $\phi_k$ of $\Phi$ associated with the $k$th brane transforms as
\[
\phi_k \to U(x^i_k) \phi_k 
\]
where $x^i_k$ is the $k$th diagonal elements of the matrix which describe the well-defined position of the $k$th brane. 

Unfortunately, our first attempt (\ref{firsttry}) cannot be right: it is straightforward to check that it is not a good representation of the gauge group since the group multiplication law is not satisfied properly. In other words, for the definition above,
\be
\label{group}
U \circ (V \circ \Phi) \ne (UV) \circ \Phi \; .
\ee
The problem comes at quadratic order in the $X$s, where we find
\[
U \circ (V \circ \Phi) - (UV) \circ \Phi = {1 \over 2} \partial_i U \partial_j V \Phi [X^j, X^i] + {\cal O} (X^3)
\]
Since this term explicitly involves a matrix commutator, we will only have a problem for noncommuting matrices. Indeed, we have already seen that (\ref{firsttry}) gives the right answer when the matrices are diagonal. 

To fix the problem, we may try to add additional terms to the gauge transformation law, with the constraint that all of these terms should involve commutators of $X$s so as not to spoil the result for diagonal matrices. It turns out that there is no way to fix the problem even at quadratic order by adding terms involving only $X^i$, $\Phi$, and the gauge transformation function $U$. On the other hand, it is not hard to check that by adding the term
\[
+ {i \over 2} \partial_i U(0) A_j(0) \Phi [X^j,X^i] 
\]
to the gauge transformation (\ref{firsttry}) that (\ref{group}) is satisfied to quadratic order. Thus, to this order, there appears to be a consistent gauge transformation law, and interestingly, it depends explicitly on the gauge field.

Proceeding order by order in $X$, one may check by hand that appropriate modifications to (\ref{firsttry}) exist such that (\ref{group}) is satisfied. Furthermore, these appear to be unique up to field redefinitions of $\Phi$ which preserve the transformation law in the abelian case. However, they are not particularly illuminating at this stage, so we will hold off discussing higher corrections until section 3, where we will obtain a systematic way to derive them. 

\subsubsection*{Gauge invariant expressions}
Once we have a consistent gauge transformation law, a natural question is how to write down gauge invariant expressions that may appear in an effective action. To obtain a singlet under the space-filling brane gauge symmetry, we might expect that any expression of the form
\be
\label{simple}
\Phi^\dagger  f(F,DF,\dots) \Phi
\ee
should suffice, where $f$ is some gauge-covariant expression built from field strengths and covariant derivatives. 

It turns out that none of these are invariant under the corrected gauge transformation we have found above. Even the simplest bilinear in $\Phi$ has a non-zero gauge variation in the case where $X$ is a matrix. However, just as for the gauge transformation law itself, we may add additional terms to these expressions to cancel the gauge variations and thereby construct gauge invariant expressions order by order in $X$. For example, we may build a singlet from a bilinear in $\Phi$ as
\bea
&&\Phi^\dagger \Phi - i[X^i, \Phi^\dagger A_i \Phi] - {1 \over 2} X^i
X^j \Phi^\dagger A_j A_i \Phi + X^i \Phi^\dagger A_i A_j \Phi X^j - {1
\over 2} \Phi^\dagger A_j A_i \Phi X^i X^j\cr
&& - {i \over 2} X^i X^j \Phi^\dagger \partial_j A_i \Phi + {i \over 2} \Phi^\dagger \partial_i A_j \Phi X^i X^j + \cdots
\label{bilin}
\eea
where we take $A$ and its derivatives to be evaluated at the origin.
As with the gauge transformation law, this expression does not look particularly enlightening. However, closer inspection reveals that it ``factors'' and may be reexpressed simply as
\[
{\cal F}^\dagger {\cal F} 
\]
where
\be
\label{covf}
{\cal F}(\Phi, A_i(y), X^i) = \Phi + i A_i \Phi X^i - {1 \over 2} A_i A_j \Phi X^j X^i + {i \over 2} \partial_j A_i \Phi X^j X^i + \dots
\ee
Evaluating the gauge variation of ${\cal F}$, we find that it has a simple transformation law
\be
\label{simple2}
{\cal F} \to U(0) {\cal F} \; .
\ee
Thus, $\cf$ is a {\it gauge covariant} version of $\Phi$. Again, one may proceed order by order and check that additional terms may be added at each order in $X$ such that $\cf$ retains the simple transformation law (\ref{simple2}). In this case, the result is not unique, but this should be expected, since for any ${\cal F}$ transforming as (\ref{simple}), we can build other objects with the same transformation law, for example
\be
\label{redefi}
{\cal F} + F_{ij}(0) {\cal F} [X^i, X^j]
\ee
Since $\cf$ is covariant, it is now simple to build invariant actions; for example, we could replace $\Phi$ with $\cf$ in any expression of the form (\ref{simple}) to obtain a gauge invariant expression. 

\subsubsection*{Summary}

To summarize, we noted that the transformation of $\Phi$ under the space-filling brane's gauge symmetry requires a rather complicated generalization of the simple rule (\ref{basic}) in the case of Abelian $X$, but that a consistent transformation law does exist. With the resulting definition, typical objects that would be gauge invariant in the Abelian case are no longer gauge invariant and must also be supplemented with additional terms. The resulting gauge invariant structures may be rewritten simply in terms of a covariant object $\cf$ with transformation rule (\ref{simple}). This is precisely the usual story: invariant actions are built from simple structures involving covariant objects.

The new field $\cf$ may be thought of as a repackaging of the degrees of freedom of $\Phi$ into an object which sits at an ordinary point, $y=0$, rather than at the matrix location $X^i$. Of course, there is nothing special about the point 0, and we may perform a translation to define
\be
\label{Fa}
{\cal F}_a (X,A(y),\Phi) = {\cal F}_0 (X-a, A(y+a),\Phi)  
\ee
such that $\cf_a$ has the transformation property
\[
{\cal F}_a \to U(a){\cal F}_a \; 
\]

Finally, we note that it is easy to write explicit expressions with the properties of ${\cal F}_a$ in the case of Abelian $X$. We simply multiply $\Phi$ by a Wilson line which connects the point $X$ where the pointlike brane is located with the point $a$,
\[
{\cal F}_a  = e^{i \int_a^X A} \; \Phi \; .
\]
The freedom of choosing different paths here is related to the possibility of redefinitions such as (\ref{redefi}) which preserve the covariant transformation law. Thus, the existence of a covariant object $\cf$ in the case of non-abelian $X$ suggests that it is possible to define a generalized Wilson line which runs from an ordinary point $a$ to a matrix location $X$!

In the next sections, we will provide a more systematic way to obtain the relationship between $\Phi$ and $\cf$ and to derive the effective actions describing intersecting brane systems. Along the way, we will discover additional interesting structures that are involved in writing invariant effective actions.

\section{Constraints from T-duality}

In this section, we will see that thinking about constraints of T-duality applied to the intersecting brane systems will allow us to simplify the problem of finding a full expression for $\cf$ and writing down the correct gauge invariant action. 

First, we note that any intersecting brane system can be obtained by
T-dualities from a system with truly pointlike branes in the presence
of branes which fill some directions $y^i$. Until section 6, we will
restrict to this latter case, ignoring all of the transverse
directions and referring to the higher dimension branes as
space-filling. We now imagine compactifying the directions described
by $y^i$ to a torus. To describe the effective action of the resulting
system, we must include additional fields describing the winding modes
of strings stretched between the pointlike branes and their images. We
will denote these degrees of freedom by $X_{\vec{n}}$ where $n^i$ are
integers and $2 \pi R n^i$ are the locations of the relevant image (we
assume the torus is rectangular with periods of length $2 \pi R$).

Given the full action for the system on a torus, it would be easy to obtain the action for the noncompact case; we essentially just ignore the winding modes. The benefit of thinking about the compactified system is that we now have an alternate picture obtained by performing T-duality on all directions. In this picture, the space-filling branes become pointlike and the pointlike branes become space-filling. Ignoring the winding modes in the original picture corresponds to choosing a constant gauge field in the dual picture. Thus, our original problem is equivalent to finding the action for pointlike branes on a torus in the presence of a constant gauge field on the space-filling branes.

This is not obviously simpler, since we must include winding modes in the dual picture. However, there is nice way to include the winding modes starting from an action without them: following Taylor \cite{wati}, to describe pointlike branes on a torus, we consider the pointlike branes plus an infinite collection of image branes in noncompact space, and then keep only the degrees of freedom which are invariant under the group of translations that define the torus. 

Thus, starting with an action for pointlike branes on a noncompact space with a constant gauge field on the space-filling brane, we can apply the quotienting prescription to include the winding modes, T-dualize to obtain the action for the pointlike branes (without winding modes) on a torus in the presence of an arbitrary gauge field, and decompactify to get an action for our original situation. If desired, we could apply the quotienting prescription again to compactify the original picture and obtain an action with all winding and momentum modes included. 

In this way, we can obtain the full action for pointlike and space-filling branes on a torus from the much simpler action describing only the zero-modes on both the torus and dual torus. In addition, the transformation laws for all the fields in each picture will follow from the transformation laws in the simplest situation where only the zero-modes are kept. Let us now see how all this works in detail. 

In section 3.1, we describe the action of T-duality on the fields of our system. In section 4, we discuss the action for the zero-modes on the torus, deriving the general action consistent with all of the symmetries and T-duality. In section 5, we review the quotienting prescription for incorporating the winding modes and thus use the zero-mode action to obtain the full action with spatially dependent fields.

\subsection{Details of T-duality}

We begin by describing how T-duality acts on the variables describing our intersecting brane system on a torus. 

Let $S_{T^d(R)}(X^i_{\vec{n}}, A_i(y), \Phi)$ denote the full action
for pointlike and space-filling branes on the original torus. In the
dual picture, we have a corresponding set of variables
$\tilde{A}_i(\tilde{y}), \tilde{X}^i_n, \tilde{\Phi}$ and a dual
action $S_{T^d(\tilde{R})}$. Here, $\tilde{R}$ is the radius of the
dual torus, which in our conventions is given by $\tilde{R} = 2 \pi /
R$ (we set $2 \pi \alpha' = 1$ throughout this paper). Since these two
actions describe the same physical system, they should be equal when
the dual variables are written in terms of the original
ones,\footnote{In fact, it is also true that the original action in
the original variables should have the same functional form as the
dual action in the dual variables, since the types of branes are the
same in both pictures. We will use this stronger statement in section
4.2.} 
\be
\label{constraint}
S_{T^d(\tilde{R})} (\tilde{X}_{\vec{n}},\tilde{A}(\tilde{y}), \tilde{\Phi}) = S_{T^d(R)}(X_{\vec{n}},A(y), \Phi)
\ee
For the gauge field and scalars, the duality transformation is well known:
\bea
\label{T1}
\tilde{A}_i(\tilde{y}) &=& \sum_n X^i_{\vec{n}} e^{-i \tilde{y} \cdot n / \tilde{R}} \cr
\tilde{X}^i_{\vec{n}} &=& {1 \over (2 \pi R)^d} \int dy A_i(y)  e^{i y \cdot \vec{n}/R}
\eea
For the bifundamental, one is tempted simply to identify $\tilde{\Phi}$ with $\Phi^\dagger$; however, this cannot be quite right. To see this, note that $\Phi$ is invariant under a translation in the original torus, which acts on the other fields as
\[
X_0^i \to X_0^i + k^i \qquad \qquad A_i(y) \to A_i(y-k)  \; .
\]
In terms of the dual variables, this becomes
\[
\tilde{A}_i \to \tilde{A}_i + k_i  \qquad \qquad \tilde{X}^i_{\vec{n}} \to e^{-i k \cdot n /R } \tilde{X}^i_{\vec{n}}
\]
In the limit $R \to 0$ where the dual torus becomes non-compact, this becomes a gauge transformation on $\tilde{A}$ corresponding to $\tilde{U} = e^{-i k \cdot \tilde{y}}$. Under such a transformation, $\tilde{\Phi}^\dagger$ certainly transforms non-trivially (as we have seen in section 2), thus it cannot be the same as $\Phi$. On the other hand, in section 2, we noticed the existence of fields $\tilde{\cf}_0$ related to $\Phi$ which are invariant under such a gauge transformation. Thus, we postulate that under T-duality, $\Phi$ is exchanged with an object $\tilde{\cf}^\dagger_0$ with the properties discussed in section 2.

More generally, we defined in (\ref{Fa}) a family of objects $\cf_a$ associated with arbitrary points $a$ on the original space. Since these are related to $\cf_0$ by translations $X_0 \to X_0 - a$, they should be exchanged with fields $\tilde{\Phi}_a$ which are the image of $\tilde{\Phi}$ under transformations $A \to A - a$. 

Thus, we propose that the fields $\Phi_a$ that transform simply under translations are exchanged with fields $\cf_a$ that transform simply under gauge transformations, 
\bea
\label{T2}
\tilde{\Phi}_a = \cf^\dagger_a \cr
\tilde{\cf}_a = \Phi^\dagger_a 
\eea
From this point forward, it will be convenient simply to define $\cf_a$ as the image of $\Phi^\dagger_a$ under T-duality, where $\Phi_a$ is the image of $\Phi$ under the transformation which maps $A \to A - a$. We will see that this definition fixes $\cf_a$ in terms of $\Phi$ up to a certain class of field redefinitions.

\section{Action and transformations for zero-modes}

Using T-duality, we have argued that the full action for pointlike and space-filling branes on a torus is completely determined from the action for the zero-modes, in which all fields on the original torus and the dual torus are taken to be spatially constant. In this section, we would like to understand the constraints on this reduced action which result from gauge invariance, translation invariance, and consistency with T-duality and to determine the relationship between $\Phi$ and $\cf$ in this simpler context.

Let $S_0(X^i, A_i, \Phi)$ be the zero-mode action we would like to determine. This action should be invariant under translations, gauge transformations on $X$, and gauge transformations on $A$. To write the transformation rules for the fields, we define, as above, $\Phi_a$ to be the image of $\Phi$ under transformations $A \to A - a$, and define $\cf_a$ to be the image of $\Phi^\dagger_a$ under T-duality.

The action should then be invariant under translations, symmetry transformations on $X$ and $A$ and constant shifts in $A$. By our definitions, it follows that the symmetry transformations act as
\bea
T_X \; &:& \qquad X^i \to X^i + k^i \qquad \Phi_a \to e^{i k \cdot a}
\Phi_a \qquad \cf_a \to \cf_{a-k} \cr
G_X \; &:& \qquad X^i \to V X^i V^{-1} \qquad \Phi \to \Phi V^{-1} \cr
G_A \; &:& \qquad A_i \to U A_i U^{-1} \qquad \Phi \to U \Phi  \cr
T_A \; &:& \qquad A_i \to A_i + k_i \qquad \cf_a \to e^{-i k \cdot a}
\cf_a \qquad \Phi_a \to \Phi_{a-k}
\label{TA}
\eea
Here, the transformation rule for $\Phi_a$ under $T_A$ follows by our definition of $\Phi_a$, while the transformation rule for $\cf_a$ under $T_X$ follows from this by T-duality. Also, $\Phi = \Phi_0$ should be invariant under translations since the p-p' strings should be affected only about the relative positions of the pointlike and space-filling branes. The general transformation rule for $\Phi_a$ under $T_X$ follows from this invariance together with the composition rule for translations and gauge transformations:
\[
\Phi_a = (e^{-ia \cdot y} \circ \Phi) \to (e^{-ia \cdot(y-k)} \circ \Phi) = e^{ia \cdot k}\Phi_a
\] 
Finally, the transformation rule for $\cf_a$ under $T_A$ follows from this by T-duality.

Note that these $T_A$ transformations correspond to physical translations on the dual
torus and are only gauge transformations when the translations are by multiples of an entire period of the torus. Even though these transformations leave the gauge field constant, they take the form of gauge transformations (\ref{gauge}) with nonconstant gauge transformation parameter $U = e^{-i k \cdot y}$. Thus, the difficulties addressed in section 2 with writing down transformation rules directly in terms of $\Phi$ remain even in this much simplified context, and it is still necessary to use $\cf$ in order to write simple transformation laws.. 

 We will now argue that these assumed transformation properties are consistent and in fact determine $\Phi_a$ and $\cf_a$ in terms of $\Phi$ up to a certain class of field redefinitions. In the section 4.2, we will then make use of the fields $\Phi_a$ and $\cf_a$ to describe the most general actions invariant under all of the symmetries and consistent with T-duality.

\subsection{Determining the covariant bifundamental $\cf$.}

In this section, we would like to determine explicitly the covariant
bifundamental $\cf_a$ (and its dual $\Phi_a$) in terms of our original
variables. Since $A$ no longer has any spatial dependence, we must now
have
\be
\label{opdef}
\cf_a \equiv {\cal O}(A, X-a) \Phi
\ee
where
\bea
\label{opexp}
{\cal O}(A, X) \Phi &=& \sum_n \sum_{\sigma_n} c_{\sigma_n}
A_{i_{\sigma(1)}} \cdots A_{i_{\sigma(n)}} \Phi X^{i_n} \cdots X^{i_1}
\cr
&=& \Phi + c_1 A_{i_1} \Phi X^{i_1} + c_{12} A_{i_1} A_{i_2} \Phi
X^{i_2} X^{i_1} + c_{21} A_{i_2} A_{i_1} \Phi X^{i_2} X^{i_1} + \dots
\eea
Here, we are making a Taylor expansion in powers of $X$ and summing
over permutations $\sigma$ which determine the order in which the $X$
indices are contracted with those of the $A$s. The coefficients
$c_{\sigma_n}$ are to be determined. Note that ${\co} \Phi$ should be
understood as a linear operator acting on $\Phi$ (rather than simply
multiplication).

Given $\cal{O}$, we can formally invert the power series to define
${\cal O}^{-1}$, and in terms of this,
\be
\label{inv}
\Phi = {\cal O}^{-1}(A, X-a) \cf_a
\ee
With the operator $\co$, the symmetry transformation rules above and
the action of T-duality may be expressed completely in terms of
$\Phi$. By demanding that the resulting transformations are
well-defined and consistent, we will now find constraints on $\co$
that determine it up to field redefinitions.

\subsubsection*{Constraints from T-duality}

A first constraint on $\co$ comes from the requirement that $\cf_a$
and $\Phi_a$ should be exchanged under T-duality. Rewriting
(\ref{opdef}) in the T-dual variables (for the case $a=0$), we have
\[
\tilde{\Phi}^\dagger = \co(\tilde{X},\tilde{A}) \tilde{\cf}^\dagger
\]
Taking the adjoint gives
\be
\label{dualrel}
\tilde{\Phi} = \co^\dagger(\tilde{A}, \tilde{X}) \tilde{\cf}
\ee
where $\co^\dagger$ takes the same form as (\ref{opexp}) but with
coefficients $c^\dagger_\sigma = c^*_{\sigma^{-1}}$. On the other
hand, since the T-dual system is equivalent (up to an exchange in the
rank of $X$ and $A$) to the original system, the relation (\ref{dualrel}) between $\tilde{\Phi}$ and
$\tilde{\cf}$ should have the same form as the relation (\ref{inv}) between
$\Phi$ and $\cf$, so we must require that
\be
\label{unitary}
\co^\dagger = \co^{-1}
\ee
Thus, given our definitions for inverse and adjoint, consistency with T-duality implies that the operator $\co$ must be unitary.

\subsubsection*{Constraints from gauge and translation invariance.}

The symmetries $T_X$, $G_X$, and $G_A$ are already expressed simply in
terms of $\Phi$. The image $\Phi_k$ of $\Phi$ under the remaining
symmetry $T_A$ may be written directly in terms of $\Phi$ using the
operator $\co$. From (\ref{TA}) and (\ref{inv}), we find
\bea
\label{phik}
\Phi_k &=& {\cal O}^{-1}(A-k, X-a) e^{i k \cdot a} \cf_a \cr
&=& e^{i k \cdot a} {\cal O}^{-1}(A-k, X-a){\cal O}(A,X-a) \Phi
\eea
Here, the final result was obtained from the transformation rule for $\cf_a$, however, it cannot depend on which $a$ we choose. Therefore, it must be that that all $a$ dependence on the right side of this expression vanishes. 

Differentiating the right-hand side of (\ref{phik}) with respect to $a$ and setting this to zero, we find
\[
0 = i k^i {\cal O}_-^{-1}  {\cal O}  \Phi + \partial_{X^i} {\cal O}_-^{-1}  {\cal O} \Phi + {\cal O}_-^{-1} \partial_{X^i} {\cal O} \Phi  \; . 
\]
Here, we have defined $\co_- \equiv \co(A-k, X-a)$ and
\be
\label{deriv}
\partial_{X^i} f(X) \equiv \partial_{y^i} f(X+y)|_{y=0} \; .
\ee
Acting with $\co_-$ on the left and using $\partial \co^{-1} = - \co^{-1} \partial \co \co^{-1}$, we obtain
\[
(\partial_{X^i} \co_- {\cal O}_-^{-1} - \partial_{X^i} \co \co^{-1} - i k_i ) \co \Phi = 0
\] 
Since $\co \Phi = \cf$ can be arbitrary, the expression in brackets should vanish as an operator relation. This provides a constraint
\be
\label{strongC}
\partial_{A_j} ( \ca_i ) = \delta_{ij}
\ee
where we have defined
\be
\label{ca}
\ca_i \equiv -i \partial_{X^i} \co \co^{-1}
\ee 
and $\partial_{A_i}$ is defined in a similar way to (\ref{deriv}). We will see a physical interpretation for $\ca$ presently. 

\subsubsection*{Interpretation of $\ca$}

Before discussing the solution to the constraints, we note that the combination $\ca_i = -i \partial_{X^i} \co \co^{-1}$ which appears plays a special role in the transformation rules: it gives the infinitesimal transformation law for $\cf_a$ under $T_A$ or $\Phi$ under $T_X$.

For example, under a $T_X$ translation $X \to X+k$, $\cf_a$ transforms as
\[
\cf'_a = \co(A, X - a + k) \co^{-1} (A, X-a) \cf_a
\]
using the fact that $\Phi$ is invariant. For an infinitesimal transformation, we find
\be
\label{ftrans2}
\delta \cf_a = k_i \partial_{X^i} \co \co^{-1} \cf_a  = i k_i \ca_i (A, X-a) \cf_a
\ee
By similar manipulations or by T-duality, the infinitesimal transformation rule for $\Phi$ under a shift $A \to A +k$ is 
\be
\label{phitrans}
\delta \Phi_a = -i k_i \ca_i^\dagger (A-a, X) \Phi_a 
\ee

The constraint equation (\ref{strongC}) implies that $\ca_i$ must be of the form
\be
\label{OAR1}
\ca_i = A_i + {\cal R}_i
\ee
where $\cR_i$ is an operator involving $A_i$ only in commutators (objects built from dimensionally reduced field strengths and covariant derivatives),
\be
\label{coms}
{\cal R}_i (A, X) \Phi = r_1 [A_j, A_i] \Phi X^j + r_2 [A_j,[A_k,A_i]] \Phi X^j X^k + r_3 [A_k,[A_j, A_i]] \Phi X^j X^k + \dots 
\ee
such that ${\cal R}_i (A+k, X) = {\cal R}_i (A, X)$.

It is interesting to check that the rules (\ref{ftrans2}) and (\ref{phitrans}) give proper representations for the Abelian group of translations. This requires that
\[
\delta_{k} (\delta_{l} \cf) - \delta_{l} (\delta_{k} \cf) = 0
\]
which will be satisfied if and only if
\be
\label{flat}
\partial_{X^i} \ca_j - \partial_{X^j}\ca_i - i[\ca_i, \ca_j] = 0 \; .
\ee
Thus, the operator $\ca_i$ which determines the infinitesimal transformation rules must coincide with $A_i$ up to commutator terms and must obey the relations of a flat connection with derivatives defined as in (\ref{deriv}). This latter equation is automatically satisfied by virtue of the definition (\ref{ca}) in terms of $\co$, which looks like the usual general expression for a flat connection, where $\co$ plays the role of the gauge transformation function.   

\subsubsection*{Solving the constraints}

We have seen that the rules above define a consistent realization of the symmetry algebra and the action of T-duality as long as $\co$ is ``unitary'', $\co^{-1} = \co^\dagger$ and the associated ``flat connection'' $\ca_i \equiv -i \partial_{X^i} \co \co^{-1}$ agrees with the constant $A_i$ up to commutator terms, 
\be
\label{con2}
\partial_{A_i} \ca_j = \delta_{ij} \; .
\ee
When either $X$ or $A$ are abelian, it is easy to see that the constraints uniquely determine 
\be
\label{abelian}
\co(A,X) = e^{i A \cdot X} 
\ee
equivalent to a Wilson line from 0 to $X$ in the case where $X$ is abelian, or Wilson line of the dual gauge field $\tilde{A} = X$ from 0 to $\tilde{X} = A$ in the case where $A$ is abelian.

In the general case, we have not been able to obtain a simple closed form expression for $\co$. However, one may apply the brute-force approach of plugging in the general expansion (\ref{opexp}) into the constraint equations to determine a set of algebraic equations satisfied by the coefficients $c_\sigma(n)$. At each order $n$, this gives a finite set of linear equations for the $c_\sigma(n)$ (as we show in appendix A), which may be solved to obtain the general solution in terms of a certain number of free parameters. An efficient method of carrying out this procedure, which may be implemented on a computer, is detailed in appendix A. Following this procedure, we have checked that solutions exist at least to fifth order in $A$. To cubic order, the general solution is
\bea
\ba{lrrrr} \co \Phi  =  &\Phi & + \; i A_i \Phi X^i & - \; ({1 \over 2} + ia) A_i A_j \Phi X^j X^i & + \; (a+ib) A_i A_j A_k \Phi X^k X^j X^i \cr
& & & + \; i a A_i A_j \Phi X^i X^j & - \; ({i \over 6} + {a \over 2} + ib +ic) A_i A_k A_j \Phi X^k X^j X^i \cr
&&&& + \; (i c -{a \over 2}) A_j A_i A_k \Phi X^k X^j X^i \cr
&&&& + \; ({i \over 6} - d - i{b \over 2}) A_j A_k A_i \Phi X^k X^j X^i \cr
&&&& + \; ({i \over 6} - i {b \over 2} + d) A_k A_i A_j \Phi X^k X^j X^i \cr
&&&& - \; ({i \over 6} - ib) A_k A_j A_i \Phi X^k X^j X^i 
\ea 
\label{leading}
\eea
where $a,b,c,$ and $d$ are arbitrary real parameters. This corresponds to an associated infinitesimal translation operator 
\[
\ca_i \Phi = A_i \Phi + {i \over 2} [A_j, A_i] \Phi X^j - {1 \over 6} [A_j,[A_k,A_i]] \Phi X^k X^j + \alpha [A_k, [A_j, A_i]] \Phi [X^k, X^j] + \cdots
\] 
where $\alpha = {b \over 2} + c - i d+i {a \over 2}$.\footnote{It is interesting to note that this general result for $\ca$ may be obtained directly from the relation (\ref{con2}) together with the consistency condition (\ref{flat}). Given the solution for $\ca$ obtained in this way, the general solution for $\co$ may then be obtained by solving (\ref{ca}), or equivalently $(\partial_{X^i} - i \ca_i) \co = 0$.}

\subsubsection*{Arbitrary Parameters and Field Redefinitions}

The general solution (\ref{leading}) to the constraint equation involves a number of arbitrary parameters, with new parameters appearing at each order in the expansion. We will now argue that these do not correspond to essentially different choices for the symmetry transformation rules, but rather to field redefinitions which preserve the stated transformation properties of $\cf_a$ and $\Phi_a$.

Consider some consistent set of symmetry transformation rules defined by an operator $\co$ satisfying the constraints above. Let $\Phi'$ be a field redefinition of $\Phi$ which is linear in $\Phi$, 
\be
\label{redef}
\Phi' = \cp (A,X) \Phi 
\ee
where $\cp$ is an operator of the form (\ref{opexp}). It is clear that $\Phi'$ will have the same transformation properties as $\Phi$ under $G_X$ and $G_A$ for any choice of $\cp$. Further, $\Phi'$ will transform in the same way as $\Phi$ under $T_X$ (i.e. trivially) as long as 
\be
\label{pconst}
\partial_{X^i} \cp = 0
\ee
In other words, $X^i$ should appear in $\cp$ only in commutators. 

The redefinition on $\Phi$ also implies a redefinition of $\cf$, the image of $\Phi$ under T-duality. To determine this, note that rewriting (\ref{redef}) in terms of the T-dual variables, we have
\beas
(\tilde{\cf}')^\dagger &=& \cp(\tilde{X},\tilde{A}) \tilde{\cf}^\dagger \cr
&=& \cp(\tilde{X},\tilde{A}) \co^\dagger(\tilde{X}, \tilde{A})  \tilde{\Phi}^\dagger \cr
&=& \cp(\tilde{X},\tilde{A}) \co^\dagger(\tilde{X}, \tilde{A}) (\cp^\dagger(\tilde{X}, \tilde{A}))^{-1} (\tilde{\Phi}')^\dagger
\eeas
Going back to the original variables, this gives an expression for $\Phi'$ in terms of $\cf'$ which may be inverted to give 
\[
\cf' = \cp^\dagger \co \cp^{-1} \Phi'
\]
where we have used $\co^\dagger = \co^{-1}$. Thus, our field redefinition leads to a transformed expression for $\co$ given by
\be
\label{oprime}
\co' = \cp^\dagger \co \cp^{-1}
\ee

We may now check that $\co'$ gives a new solution to the constraint equations assuming that $\co$ does. For the unitarity constraint (\ref{unitary}), we have\footnote{Note that with our definition of adjoint, taking the adjoint of a product of operators does not reverse the order of the operators.}
\[
(\co')^\dagger = \cp \co^\dagger (\cp^{-1})^\dagger = \cp \co^{-1} (\cp^\dagger)^{-1} =(\co')^{-1}
\]
To check that (\ref{strongC}) is satisfied, note that with $\partial_{X^i} \cp = 0$,
\be
\label{likegauge}
{\ca'}_i = \cp^{\dagger} \ca_i (\cp^\dagger)^{-1} -i \partial_{X^i} \cp^{\dagger}(\cp^\dagger)^{-1}
\ee
so amusingly, the field redefinition looks like a gauge transformation on our flat connection. From this expression, we find that 
\[
\partial_{A_i} \ca'_j = \delta_{ij} 
\]
using the fact that $\partial_{X^i} \cp = 0 \Rightarrow \partial_{A_i} \cp^\dagger = 0$. Thus, $\co'$ gives new solution to the constraints. We now propose that the arbitrary parameters appearing in the general solution simply reflect this ability to generate new solutions by making field redefinitions on $\Phi$. 

As a check, we may explicitly determine the effects of an arbitrary field redefinition to leading orders in $A$. To third order, the most general field redefinition of the form (\ref{redef}) satisfying (\ref{pconst}) is
\[
\Phi' = \Phi + u A_i A_j \Phi [X^i, X^j] + v A_i A_j A_k \Phi [X^i,[X^j,X^k]] + w A_i A_j A_k \Phi [X^j, [X^i,X^k]] + \cdots
\]
Inserting this expression into (\ref{oprime}), we find that the resulting expression for $\co'$ again takes the form (\ref{leading}), with 
\beas
a' &=& a + 2 Im(u)\\
b' &=& b + 2 Im(v)\\
c' &=& c + 2 Im(w) + Re(u)\\
d' &=& d + Re(v) + 2 Re(w)
\eeas
Thus, at least to this order, we can set the free parameters in the
general solution to arbitrary values using field redefinitions. It
seems likely that this will hold at higher orders also, though we do
not have a general proof. If true, this implies that up to field
redefinitions, there is a unique consistent realization of the
symmetries on the fields in the problem.\footnote{It is possible  that
there exists a canonical choice for which the expression for $\co$
takes a particularly simple form or satisfies some further natural
constraints.}   

\subsubsection*{Summary}

We have seen that our assumed symmetry and T-duality transformation rules are consistent if $\cf$ is related to $\Phi$ by (\ref{opdef}), where $\co$ satisfies the constraints 
\[
\co^\dagger = \co^{-1} \qquad \partial_{A_i} \ca_j = \delta_{ij} \qquad \ca_i \equiv -i \partial_{X^i} \co \co^{-1} \; . 
\]
When either $A$ or $X$ is abelian, these are solved uniquely by 
\[
\co = e^{i A \cdot X} \; ,
\]
while in the general case, we have found that the constraints have a unique solution (to the order we have checked) up to field redefinitions of the form
\[
\Phi' = \cp(A,X) \Phi \qquad \qquad \partial_{X^i} \cp = 0
\]
which generate new solutions for $\co$ and $\ca$ given by
\[
\co' = \cp^\dagger \co \cp^{-1} \qquad \qquad {\ca'}_i = \cp^{\dagger} \ca_i (\cp^\dagger)^{-1} -i \partial_{X^i} \cp^{\dagger}(\cp^\dagger)^{-1} \; .
\]
The leading terms in the general solution for $\co$ are given by (\ref{leading}). In the rest of the paper, we will assume that a full solution for $\co$ exists, and proceed to construct invariant actions with $\co$ as an ingredient. 

\subsection{Invariant zero-mode action}

Armed with the symmetry transformation rules defined above, we are now in a position to understand the necessary structure of an action describing the zero-modes in our Dp-Dp' system. In general, the terms not involving the bifundamental field are the same as those in the individual actions for the Dp-branes or Dp'-branes, so our interest is only in the terms containing $\Phi$. This bifundamental action should have the following properties:

\begin{enumerate}
\item
Symmetry: The action must be invariant under the symmetries $G_A$, $G_X$, $T_A$, and $T_X$.
\item
T-duality: The action should be mapped to another action of the same form under a full T-duality transformation which exchanges the pointlike and space-filling branes.  
\item
Localization/Cluster decomposition: In the case where $X$ is diagonal, the action describes $N$ pointlike branes with well-defined positions, and should split into a sum of terms describing to the individual branes, with each term reducing to the known action for a single D-brane. More generally, when $X$ is block-diagonal with the separations between individual blocks taken to infinity, the physics must be that of independent systems, each described by our original action. This constraint is simplest for the bilinear terms, where it holds classically. On the other hand, as we explain below, for terms quartic in the bifundamentals, this property does not hold classically, but only at the level of the low-energy effective action.    
\item
Supersymmetry: In $Dp-D(p+4)$ systems, the action should be strongly constrained by the requirement of invariance under eight supercharges.  
\item
Agreement with known results: The action should reproduce the known leading terms in the $\alpha'$ expansion. For the Dp-D(p+4) system, this is the dimensional reduction to p+1 dimensions of six-dimensional ${\cal N} = 1$ gauge theory with hypermultiplets in the adjoints of $U(N)$ and $U(k)$ and one bifundamental hypermultiplet, where $N$ and $k$ are the number of Dp-branes and D(p+4)-branes. 

\end{enumerate}
In the rest of this section, we discuss the constraints that symmetries, T-duality and locality impose on the action. It would be interesting to understand how to combine these constraints with the constraints associated with supersymmetry,  however we leave this for future work. We leave a discussion of specific examples to section 6.1 where we will discuss the supersymmetric Dp-D(p+4) system. 

\subsubsection{Symmetry constraints}
 
The symmetry transformation rules for the various fields were given at the beginning of section 4. To ensure $G_A$ and $G_X$ invariance, we simply require that each term in the action is a product of traces over appropriately multiplied matrices. At weak string coupling, the leading terms will involve only a single trace, coming from worldsheet disk diagrams. Below, we will consider only single-trace actions, though most of the considerations extend to multi-trace actions in an obvious way. 

To ensure invariance under $T_A$, we should build the action using the covariant bifundamental $\cf_a$ (with equal numbers of $\cf^\dagger_a$s and $\cf_a$s) and ensure that $A$ appears only in commutators. For example, bilinear terms will be invariant if $A$ and $\Phi$ appear only in combinations 
\[
B[{\cal C},a] = \cf_a^\dagger {\cal C}(A) \cf_a
\]
where ${\cal C}$ is a commutator expression. 

To understand how to ensure $T_X$ invariance, note that under a transformation $X \to X + k$, the $T_A$ invariant block just defined (or more general expressions where ${\cal C}$ is not necessarily a commutator) transforms as
\[
B[{\cal C},a] \rightarrow B[{\cal C},a-k]
\]
In other words, any such expression transforms like a field evaluated at the point $a$ on the space-filling brane. This suggests that as for ordinary fields, invariant expressions should be obtained by integrating over the whole space-filling brane. Before doing this, we note that $X-a$ (considered as a function of $a$) transforms in the same way as $B$ under $T_X$ so that any expression of the form
\[
T(a) = \sum \cf_a^\dagger f_1(A) \cf_a h_1(X-a) \cdots \cf_a^\dagger f_n (A) \cf_a h_n(X-a)
\]
will transform as a field under $T_X$ translations and be invariant under $T_A$ as long as 
\be
\label{cona2}
\partial_{A_i} T(a) = 0 \; .
\ee
For bilinear terms, the most general such expression is a sum of terms
of the form
\[
T(a) = \cf_a^\dagger {\cal C}(A) \cf_a h(X-a)
\]
where ${\cal C}$ is a commutator expression, while for higher order terms, we do not need to require that the individual functions $f_i$ are commutators, since $A$ may appear in commutators such as $[A_i, \cf_a h(X-a) \cf_a^\dagger]$.

Given some choice of $T(a)$, we may integrate in $a$ over the space-filling brane to obtain an action
\be
\label{gen1}
\int da \tr(T(a)) 
\ee
invariant under all of the symmetries. Even more generally, we could consider single-trace actions involving multiple integrals, for example
\be
\label{gen2}
\int da_1 \cdots da_n \;  K(a_i - a_j) \; \tr(T_1(a_1) \cdots T_m(a_n)) 
\ee
where $K$ is an arbitrary function of the differences. In this case, the individual $T_i$'s need not satisfy the constraint (\ref{cona2}) as long as the entire expression in the trace does. While these terms look nonlocal, it should be remembered that so far, none of the basic fields have any spatial dependence, so there is no reason to rule them out at this point.

Before considering the remaining constraints, we note that our
discussion above has been rather asymmetrical with respect to $T_A$
and $T_X$. We have enforced $T_A$ invariance by constructing invariant
blocks out of covariant combinations of fields, while $T_X$ invariance
was enforced by integrating field-like objects over space. On the
other hand, the symmetry transformation rules are unchanged if we make
the T-duality substitutions $A \leftrightarrow X$, $\cf_a
\leftrightarrow (\Phi_a)^\dagger$. In fact, the whole discussion above
would proceed in an identical manner if we made these replacements, leading
to expressions with $T_X$ covariant objects built from $\Phi_a$ and
$X$-commutators multiplied by arbitrary functions of $A-a$ and
integrated over space. In the end, we could rewrite these invariants
by expressing $\Phi_a$ in terms of $\cf_a$, and (as we show below) the result will be a sum of terms of the form we have described
above, rather than new invariant structures. However, in writing
actions consistent with T-duality, it will
be convenient to include both terms written using $\Phi_a$ and
terms written using $\cf_a$.   

Thus, for example, single-trace invariant actions bilinear in $\Phi$ may be written either as
\be
\label{fact}
\int da  \tr(\cf^\dagger_a {\cal C}(A) \cf_a h(X-a)) 
\ee
or as
\be
\label{pact}
\int da \tr(\Phi_a {\cal C}(X) \Phi^\dagger_a h(A-a)) 
\ee
where $h$ is an arbitrary function and ${\cal C}$ is a function involving only commutators ($\partial_{X^i} {\cal C} = 0$ in our earlier notation).

\subsubsection{T-duality constraints} 

After a T-duality transformation, $N$ pointlike and $k$ space-filling branes will be mapped to $k$ pointlike and $N$ space-filling branes, so the action in the new variables should be identical to our original action up to the exchange of $N$ and $k$ and appropriate transformations on the volume of the torus and the coupling. Explicitly, if we write the action in the form $S_{\{V,g_s,N,k\}}(X,A,\cf_a)$ as above, the T-duality constraint requires  
\be
\label{Tconst}
S_{\{V,g_s,N,k\}}(\tilde{A},\tilde{X},\tilde{\Phi}^\dagger_a) = S_{\{\tilde{V},\tilde{g}_s,k,N\}}(\tilde{X},\tilde{A},\tilde{\cf}_a)
\ee
In fact it is easy to obtain the general invariant action which satisfies this constraint. We must have
\[
S = S'_{\{V,g_s,N,k\}}(X,A,\cf_a) + S'_{\{\tilde{V},\tilde{g}_s,k,N\}}(A,X,\Phi^\dagger_a) 
\]
where $S'$ is any invariant action constructed in the previous section. It is obvious that any such action satisfies the constraint. Conversely, any action $S$ satisfying the constraint may be written in this form by taking $S' = S/2$.  

As an example, a general symmetry-invariant term bilinear in $\Phi$ and consistent with T-duality should take the form
\be
\label{tact}
\int da  \tr(\cf^\dagger_a {\cal C}(A) \cf_a h(X-a)) + \int da \tr(\Phi_a {\cal C}(X) \Phi^\dagger_a h(A-a)) 
\ee
where, as above, ${\cal C}$ is a commutator term and $h$ is an arbitrary function.

\subsubsection{Locality constraints}

We would now like to consider additional constraints placed on our actions by physical considerations related to locality. We will discuss first the bilinear terms (\ref{tact}) and then consider more general terms with additional bifundamentals.

Consider the situation where $X = diag(x_1, \dots,x_n)$. In this case, the $\Phi_a$ terms in (\ref{tact}) reduce to
\[
\sum_n  \int da (\Phi_a)_n {\cal C}(0) (\Phi^\dagger_a)_n h(A-a)
\]
while the $\cf_a$ terms reduce to a sum of terms
\[
\sum_n  \int da (\cf_a^\dagger)_n {\cal C}(A) (\cf_a)_n h(x_n-a)
\]
Further, for commuting $X$, we have 
\[
\co = e^{i A \cdot X} \qquad \Rightarrow \qquad (\cf_a)_n = e^{i A \cdot (x_n - a)} \phi_n \qquad (\Phi_a)_n = e^{i x_n \cdot a} \phi_n  
\]
so that the action (\ref{tact}) becomes
\be
\label{abact}
\sum_n \left( \sum_m {i^m \over m!} h^{(i_1 \cdots i_m)} \phi^\dagger_n [A_{i_1}, \cdots [A_{i_m}, {\cal C}(A)] \cdots ] \phi_n  \; \; + \; \;  \phi^\dagger_n {\cal C}_h(A) \phi_n {\cal C}(0) \right)
\ee
where we have defined
\[
h^{(i_1 \cdots i_m)} = \int da \;  a^{i_1} \cdots a^{i_m} \; h(-a) \qquad \qquad {\cal C}_h(A) = \int da \; h(A-a) \; .
\]
We have used the notation ${\cal C}_h(A)$ since by its definition it follows that $\partial_{A^i} {\cal C}_h = 0$, so this must be a commutator expression. For these quantities to be defined as a power series expansion, the function $h$ must vanish sufficiently rapidly at infinity so that all of its multipole moments are finite, though more general functions are possible.  

Since the expression (\ref{abact}) is independent of $x^i_n$, we must demand that the individual terms in the sum over $n$ are precisely the bilinear terms in the action for a single pointlike brane, since we can imagine taking $x^i_n$ arbitrarily far apart so that the individual branes don't interact. While the single-brane action certainly has terms at all orders in the $\alpha'$ expansion, like the expression (\ref{abact}) for general $h$, it is often useful to consider only the leading terms which dominate at low energies.\footnote{Indeed, we implicitly considered this limit in section 2 by assuming that the pointlike branes had delta-function support.} Thus, it is interesting to ask whether there is an invariant action for non-abelian $X$ which reduces only to these specific leading terms, which should be a finite sum of terms of the form (taking into account T-duality)
\be
\label{specific}
S_\alpha = \phi^\dagger {\cal C_\alpha(A)} \phi h_\alpha + \phi^\dagger h_\alpha \phi {\cal C}_\alpha(0) \; .
\ee

In order that the sum over $m$ in (\ref{abact}) should give a finite set of terms, it must be that $h(a)$ has a finite number of non-zero multipole moments, and is therefore a distribution rather than an ordinary function. Physically, this is what we expect: in the $\alpha' \to 0 $ limit, individual D-branes are pointlike in the transverse directions, so the action for the bifundamentals written in the form (\ref{fact}) as an integral over the transverse space should localize to the positions of the individual branes for diagonal $X$. This will be the case if the support of $h(X-a)$ (as a function of $a$) for diagonal $X$ is the collection of diagonal elements of $X$. 

In fact, we may obtain an invariant action that reduces to a specific individual term (\ref{specific}) by choosing
\[
{\cal C} = {\cal C}_\alpha \qquad \qquad h(X-a) = h_\alpha \delta(X-a)
\]
 in (\ref{tact}), where $\delta(X-a)$ is any function that satisfies 
\[
\int da \; \delta(X-a) = \identity
\]
for any X and reduces to a diagonal matrix $diag(\delta(x_1-a), \dots ,\delta(x_n-a))$ of ordinary delta functions when evaluated on a diagonal matrix $X = diag(x_1, \dots, x_n)$. 

There are many possible definitions of a matrix delta function satisfying these properties. One choice is the symmetrized delta function 
\be
\label{deltasym}
\delta_{sym}(X-y) = \int d^9 k \; e^{ik \cdot (X-y)} \; . 
\ee
With this definition, the matrix delta function associates with any ordinary function a corresponding matrix function with symmetrized ordering prescription 
\be
\label{fsym}
\int dy f(y) \delta_{sym}(X-y) = f_{sym}(X)
\ee
This particular choice appears already in the low-energy couplings of multiple D-branes to bulk fields. For example, the leading order coupling between D0-branes and the time component of the RR one-form field may be written \cite{tv0}
\[
\int dt \int d^9y C_0(y,t) \tr(\delta_{sym}(X(t)-y))
\]
More general choices for the $\delta$ function (or more general distributions) correspond to different results for the right-hand side of (\ref{fsym}). However, as we will show below, it turns out that the most general invariant actions may be constructed by restricting to\footnote{As we discuss in section 5.3, a more canonical choice might be to symmetrize the individual commutators in ${\cal C}$ with the powers of $k \cdot X$ appearing in the definition of the matrix $\delta$-function here. This type of symmetrized structure appears in all of the leading order couplings of pointlike D-branes to bulk fields \cite{tv0}.} 
\be
\label{delcon}
h(X-a) = {\cal C}_h(X) \delta (X-a)
\ee
for some choice of $\delta$ function, which we may take to be $\delta_{sym}$. Any invariant term constructed using an alternate choice for $\delta$ or some more general distribution for $h$ may be rewritten as a sum of invariant expressions constructed from terms with $h$ satisfying (\ref{delcon}). 
 
Before discussing terms with additional bifundamentals, we note that for block diagonal $X$, the bilinear terms automatically reduce to a sum of similar terms corresponding to the individual blocks. Thus, the degrees of freedom associated with individual blocks corresponding to widely separated branes are decoupled and therefore satisfy the cluster decomposition property. 

We now consider more general terms of the type (\ref{gen1},\ref{gen2}) (or the T-dual terms involving $\Phi_a$) with four or more bifundamental fields. We first note that for the bilinear terms, the constraint that $h(X-a)$ have support only at the locations of the individual branes for diagonal $X$ implies that the action (\ref{fact}) in this limit only involves the covariant quantity $\cf_y$ for $y$ equal to one of the brane locations $x_n$.\footnote{When we add spatial dependence to the gauge field, this constraint will also imply that the the system of pointlike branes couples to the gauge field only at the position of these branes.} It seems reasonable to expect that a similar locality constraint should hold for the terms with additional bifundamentals. In the most general actions (\ref{gen2}), this constraint will be satisfied as long as one of the functions $h_m(X-a_i)$ for each $T_i$ has support only at points $a_i = x_n$ for $X = diag(x_1, \dots , x_n)$. Again, without loss of generality, we may assume that this $h_m$ takes the form (\ref{delcon}). 

It is much less obvious how to impose the cluster-decomposition property for the terms with more than two bifundamentals. For these more general terms, we can no longer demand that the classical action reduces to a sum of individual actions for diagonal or block diagonal $X$. To see this, consider the low-energy effective action for the Dp-D(p+4) system with D(p+4)-brane fields set to zero. This is the dimensional reduction to $p+1$ dimensions of {\cal N}=1 gauge theory in 6 dimensions with one hypermultiplet in the adjoint and $k$ hypermultiplets in the fundamental of $U(N)$. In particular, this action includes quartic terms in the fundamental scalars which schematically take the form
\be
\label{quartic}
\sum_{n,m} \phi^\dagger_n \phi_n \phi^\dagger_m \phi_m
\ee
where $n$ and $m$ are indices corresponding to the individual Dp-branes. These terms are independent of the locations of the Dp-branes, and appear to couple the fundamentals associated to different Dp-branes with unchanging strength even when the branes are separated arbitrarily far apart. This seems to be a clear violation of cluster decomposition, and naively indicates that this action does not capture correctly the physics of the branes. 

The resolution is that cluster decomposition is restored in the low-energy effective action obtained from integrating out open-string modes which become very massive when the Dp-branes are widely separated.\footnote{This is in contrast to simple systems of widely separated Dp-branes where it is essential that terms obtained from integrating out massive off-diagonal modes exactly cancel for static parallel configurations.} Schematically, if $X^i$ are the matrices describing transverse configurations of the Dp-branes, the effective action has terms of the form
\[
\Phi^\dagger [X,X] \Phi
\]
which give rise to couplings
\[
\phi^\dagger_n (x_n - x_m) y_{mn} \phi_m
\]
where $x_m$ and $x_n$ are the diagonal elements of $X$ which describe the positions of the Dp-branes and $y_{mn}$ are scalars arising from strings stretched between the $m$th and $n$th brane. When $x_m - x_n$ is large, $y_{mn}$ develops a large mass
\[
(x_m-x_n)^2 y^\dagger_{mn} y_{mn} 
\]
coming from $[X,X]^2$ terms so that integrating out the $y$'s gives new terms of the form
\[
\sum_{m \ne n} \phi^\dagger_n \phi_n \phi^\dagger_m \phi_m
\] 
which turn out to cancel the terms in (\ref{quartic}) with $m \ne n$. This leaves only a sum of terms
\[
\sum_n \phi^\dagger_n \phi_n \phi^\dagger_n \phi_n \; , 
\] 
corresponding to the individual branes, in accordance with the cluster decomposition principle. 

Requiring this cancellation (and thereby cluster decomposition) essentially determines the structure of the quartic bifundamental terms in the low-energy effective action from the bilinear terms. It is likely that quartic bifundamental terms are constrained in a similar way in the more general symmetry-invariant actions we are considering here, however we leave a precise determination of the form of such constraints for future work.

\subsubsection{Minimal basis of invariant actions}

We have provided a general construction of invariant actions consistent with T-duality and the property of localization for diagonal $X$. We will now see that our description includes all possible actions with these properties and describe a minimal basis of such actions.

Consider any invariant expression built from $X$, $A$, $\cf_a$, and $\Phi_a$. Using the definitions of $\cf_a$ and $\Phi_a$ in terms of $\Phi$, any such expression may be rewritten in terms of $\Phi$, $X$, and $A$ alone, and in general will take the form of an infinite power series in $X$ and $A$. Consider the terms in such an expansion with the fewest powers of $X$. For $G_X$ and $G_A$ invariance, these must take the form
\be
\label{leadX}
S_0(\Phi, X, A) = \sum \tr(\Phi^\dagger f_1(A) \Phi g_1(X) \cdots \Phi^\dagger f_n(A) \Phi g_n(X)) \; .
\ee
Further, since $\Phi$ and $A$ are invariant under $T_X$, it must be that $\partial_{X^i} S_0 = 0$, that is, we must be able to write $S_0$ with $X$ appearing only in commutators. Under infinitesimal $T_A$ transformations $A \to A + k$, $\Phi$ has a complicated transformation law given in (\ref{phitrans}); however, we note that the leading term in the variation is first order in $X$. Thus, under a $T_A$ variation, terms coming from the variation of $\Phi$ must cancel against the variation of $A$ in terms at higher orders in $X$, but the terms coming from the variation of $A$ must cancel on their own. This requires that $\partial_{A_i} S_0 = 0$; in other words, we must be able to write $S_0$ with $A$ appearing only in commutators. 

Thus, any invariant expression must have leading terms $S_0$ in $X$ of the form (\ref{leadX}) which satisfy 
\be
\label{cleading}
\partial_{X^i} S_0 = \partial_{A_i} S_0 = 0 \; .
\ee
This will be true if all $f_i(A)$ and $g_i(X)$ in (\ref{leadX}) are commutator expressions, but there are more general possibilities (at quartic and higher order in $\Phi$) such as
\be
\label{freak}
\tr(\Phi^\dagger A_i \Phi X^i \Phi^\dagger \Phi - \Phi^\dagger \Phi X^i \Phi^\dagger A_i \Phi)  
\ee
which may be written either with $X$ appearing only in commutators or $A$ appearing only in commutators, but not both simultaneously.

We next note that by the constructions of the previous section, any expression of the form (\ref{leadX}) satisfying (\ref{cleading}) may be completed into an invariant action. Specifically, for 
\be
\label{snought}
S_0 = \tr(T(\Phi,A,X))
\ee
we may choose an invariant completion
\be
\label{complet}
S = \int da \; \tr(T(\cf_a, A, X-a) \; \delta(X-a)) \; .
\ee
This is invariant by the arguments of the previous section since $A$ appears in $T$ only in commutators. To see that it reduces to $S_0$ at leading order in $X$, note first that $T(\cf_a,A,X-a) = T(\cf_a,A,X)$ since we required that $T$ may be written with $X$ appearing only in commutators. Then, in the expansion of $T(\cf_a,A,X)$ in terms of $\Phi$, $A$, and $X$, we will have a leading term $T(\Phi,A,X)$ with higher order terms all having either additional factors of $X$ or additional factors of $a$, from the expansion of $\cf_a$s. The integral over $a$ will give an additional factor of $X$ for each factor of $a$, since
\[
\int da \; a^{i_1} \cdots a^{i_n} \; \delta(X-a) = X^{(i_1} \cdots X^{i_n)} \; , 
\]
so the lowest order piece in the final expression will be exactly (\ref{snought}).   

Finally, we show that given a minimal basis $\{S^\alpha_0 \}$ of expressions (\ref{leadX}) satisfying (\ref{cleading}), any invariant action $I$ may be written uniquely as a linear combination of terms given by the invariant completions  (\ref{complet}) of this minimal basis, which we denote by $\{S^\alpha \}$.\footnote{Here, we are assuming that the invariant expression admits an expansion in powers of the fields.} To see this, note that by assumption, the leading terms in $I$ may be written uniquely as a linear combination of terms in $\{S^\alpha_0 \}$. Subtracting from $I$ the invariant completion of this linear combination, we obtain a new invariant action with leading terms at higher order in $X$. Repeating this procedure at each order in $X$ will eventually lead to an expression for $I$ as a linear combination of terms in $\{S^\alpha\}$. This answer is unique since the choice of terms in $\{S^\alpha_0 \}$ is unique by our assumption that this is a minimal basis.   

Thus, we have shown that any term (\ref{leadX}) satisfying (\ref{cleading}) has an invariant completion given by (\ref{complet}) and that the invariant completions of a minimal basis of such terms gives a minimal basis of invariant actions. We note in particular that considering more general choices for the definition of $\delta(X-a)$ or more general actions (\ref{gen2}) involving multiple integrals will not lead to any new invariant structures. This also proves our earlier assertion that the T-dual invariant expressions written in terms of $\Phi_a$ may always be rewritten in terms of invariant expressions involving $\cf_a$. On the other hand, we could have written the minimal basis completely in terms of $\Phi_a$ by choosing the completion of $S_0$ to be
\be
\label{complet2}
S = \int da \; \tr(T(\Phi_a^\dagger, X, A-a) \; \delta(A-a)) \; .
\ee

We can similarly characterize the complete set of actions which are both invariant under the symmetries and consistent with T-duality. In this case, terms at lowest order in $X$ will either have fewer powers of $X$ than $A$ or have an equal number and satisfy
\[
S_0(\Phi, X, A) = S_0(\Phi^\dagger, A, X) 
\]
(terms with more $A$s than $X$s are T-dual at leading order to terms with more $X$'s than $A$s). Conversely, given any such leading terms, written as 
\[
S_0 = \tr(T(\Phi,X,A)) 
\]
for fewer $X$'s than $A$s or
\[
S_0 = \tr(T(\Phi,X,A)) + \tr(T(\Phi^\dagger,A,X))
\]
for equal $X$'s and $A$'s, we may write an invariant completion consistent with T-duality as
\[
S = \int da \; \tr(T(\cf_a, A, X-a) \; \delta(X-a)) + \int da \; \tr(T(\Phi^\dagger_a, X, A-a) \; \delta(A-a))
\]
By the same arguments as above, the set of these completions for some minimal basis of $S_0$s will provide a minimal basis of invariant actions consistent with T-duality. 

As an example, we may write a general invariant bilinear action consistent with T-duality as a sum of terms of the form
\be
\label{bigen}
\int da \tr( \cf_a^\dagger {\cal C}_1 (A) \cf_a {\cal C}_2 (X) \delta(X-a) ) + \int da \tr( \Phi_a {\cal C}_1(X) \Phi^\dagger_a {\cal C}_2(A) \delta(A-a)) 
\ee
where ${\cal C}_1$ and ${\cal C}_2$ are both commutator expressions.

\subsubsection{Examples}  

In this section, we look at some explicit examples of the higher order terms required to make leading order expressions consistent with symmetries and T-duality. 

Particularly interesting are leading terms in the action with no $A$ dependence, since we will see that symmetries and T-duality predict an infinite series of couplings to the gauge field associated with any such terms. We first consider bilinear terms in the $A=0$ action, which must take the form
\be
\label{leado}
\tr(\Phi^\dagger \Phi {\cal C}(X)) \; , 
\ee
where ${\cal C}$ is a commutator expression. From the discussion at the end of the previous subsection, the symmetry-invariant completion of these terms consistent with T-duality must be
\be
\label{examp}
\int da \tr( \cf_a^\dagger \cf_a {\cal C} (X) \delta(X-a) ) + \int da \tr( \Phi_a \Phi^\dagger_a {\cal C}(A) \delta(A-a)) 
\ee
up to invariant expressions of the form (\ref{bigen}) with higher order leading terms. Including the leading higher order invariant expressions with arbitrary coefficients and expanding everything out in terms of $\Phi$ using (\ref{opdef}), (\ref{phik}), and (\ref{leading}), we find that up to second order in $X$ (beyond the leading terms) the most general invariant action built from (\ref{leado}) and consistent with T-duality is
\be
\ba{llll} \tr ( & \Phi^\dagger \Phi \;  {\cal C}(X) &+ i \Phi^\dagger A_i \Phi \; [X^i,{\cal C}(X)] &+ ({1 \over 2}-\alpha)  \Phi^\dagger {\cal C}(A) A_i  A_j \Phi \; [X_i,X_j] \cr  
& +\Phi^\dagger {\cal C}(A) \Phi & &+ {1 \over 2} \Phi^\dagger A_i {\cal C}(A) A_j \Phi \; [X_i,X_j]  \cr
&&& + (\alpha + \beta) \Phi^\dagger A_i A_j {\cal C}(A) \Phi \; [X_i,X_j]  \cr
&&& - ({1 \over 4} - \alpha) \Phi^\dagger A_i A_j \Phi \; [X^i, [X^j, {\cal C}(X)]] \cr
&&& - ({1 \over 4} + \alpha) \Phi^\dagger A_j A_i \Phi \; [X^i, [X^j, {\cal C}(X)]] \cr
&&& + ({3 \over 4} + i a) \Phi^\dagger A_i A_j \Phi \; {\cal C}(X) [X^i, X^j]  \cr 
&&& + ({1 \over 4} - i a + \beta) \Phi^\dagger A_i A_j \Phi \; [X^i, X^j] {\cal C}(X) 
+ \dots )\ea 
\label{biglong}
\ee 
Here, $a$ is the arbitrary coefficient appearing in the expression (\ref{leading}) for $\co$ (related to the freedom of performing field redefinitions) while $\alpha$ and $\beta$ are the arbitrary coefficients of the leading higher order invariants, corresponding to expressions (\ref{bigen}) with $({\cal C}_1, {\cal C}_2) = ([A_i,A_j], [X^i,[X^j,{\cal C}]])$ and $({\cal C}_1, {\cal C}_2) = ([A_i,A_j]/2, [X^i,X^j] {\cal C})$ respectively. 

Note that all terms in (\ref{biglong}) are expressed in terms of $X$ commutators, as required by $T_X$ invariance.    Further, we see that $T_A$ invariance together with T-duality fix uniquely the linear coupling to the gauge field, yielding a current 
\[
J^i = i \Phi [X^i, {\cal C}(X) ]\Phi^\dagger \; .
\]
At quadratic order in $A$, the seven terms allowed by $T_X$ invariance
are fixed up to two arbitrary coefficients given some choice of the
field redefinition parameter $a$. Thus, while the higher order terms are not completely determined by the symmetries, they are strongly constrained.

As a special case, we note that for ${\cal C} = 1$, the general invariant expression (\ref{biglong}) reduces to 
\[
\tr( \Phi^\dagger \Phi + (1 + \beta) \Phi^\dagger [A_i, A_j] \Phi [X^i, X^j] + \dots)
\]
so that with the choice $\beta = -1$, both the linear and quadratic terms vanish. In fact, $\tr(\Phi^\dagger \Phi)$ represents an invariant action without any additional terms, because of the relation
\[
\tr(\cf_a^\dagger \cf_a) = \tr (((\co(A,X-a)\Phi)^\dagger (\co(A,X-a) \Phi) = \tr(\Phi^\dagger (\co^\dagger \co \Phi)) = \tr (\Phi^\dagger \Phi) \; .
\] 
Here, we see that because of the unitarity property of $\co$, the simple bilinear $\tr(\Phi^\dagger \Phi)$ is not only invariant under $T_X$, $G_X$, and $G_A$, but also $T_A$, since it is equivalent to the invariant expression on the left. Furthermore, this expression alone retains its form under T-duality (by the equivalence with the expression on the left for $a=0$) and trivially obeys the localization and cluster-decomposition properties. Thus, an arbitrary mass term
\be
\label{simplest}
m^2 \tr(\Phi^\dagger \Phi)
\ee
satisfies all of the constraints we have discussed. This is only true in the case where we restrict to constant gauge fields. Looking back at the invariant expression (\ref{bilin}) for spatially varying gauge fields, we may check that taking the trace yields terms quadratic in $X$ depending on derivatives of the gauge field.

Finally, we consider the simplest quartic terms 
\be
\label{simpquart}
\tr( \Phi^\dagger \Phi  \Phi^\dagger \Phi)
\ee
since these will appear below in our discussion of the Dp-D(p+4) system. These will be completed to an invariant expression
\[
{1 \over 2} \int da \tr(\cf_a^\dagger \cf_a \cf_a^\dagger \cf_a \delta(X-a)) + {1 \over 2}  \int da \tr(\Phi_a^\dagger \Phi_a \Phi_a^\dagger \Phi_a \delta(X-a))
\]
up to higher order invariants. As for the bilinear terms, the first arbitrary coefficients appear at second order in $X$ and $A$, so the linear coupling to the gauge field is uniquely determined to be\footnote{Note that while there is a higher order invariant beginning with the expression (\ref{freak}) linear in $X$ and $A$, this term maps to negative of itself under T-duality and thus cannot appear in an action consistent with T-duality.}  
\[
\tr( \Phi^\dagger \Phi  \Phi^\dagger \Phi + 2i \Phi^\dagger A_i \Phi [X^i, \Phi^\dagger \Phi]  + \dots) \; .
\]

We emphasize that all results in this section have assumed that the gauge field $A$ is spatially constant. However, as we have argued previously, knowledge of the invariant actions for the zero-modes will completely determine the possible invariant actions in the more general cases where the fields have spatial dependence. We will see how this works in detail presently.

\section{Adding the winding models}

In the previous section, we have determined the general structure of
the action describing the zero-modes of pointlike and space-filling
branes on a torus, with all winding modes (or equivalently, spatial
dependence) ignored. In this section, we follow the general procedure
of putting D-branes on a quotient space to add the winding modes and
thus obtain invariant actions with a spatially dependent gauge field on a
noncompact space or for pointlike and space-filling
branes on a torus with all modes included. For the reader wishing to skip the details of the derivation, our results are summarized and discussed in section 5.3. 

Let us briefly recall the general procedure for obtaining the action
for D-branes on a quotient space \cite{dm}.
We start with a set of variables $\chi$ describing the branes and their images on the covering space and an action $S: \chi \to R$, invariant under some symmetry group G. We then keep only the field configurations which are invariant under some subgroup $H \in G$ and parameterize these configurations by a reduced set of variables $\chi_H$. The action for the quotient is then $S(\chi(\chi_H))$ and the symmetry group of this new action will be the subgroup of transformations $G_H \in G$ which map the set of  $H$-invariant field configurations into itself (and thus give a well defined action on the new variables $\chi_H$).

In the case at hand, our starting point is any invariant zero-mode action from the previous section, 
\[
S(X^i, A_i, \Phi)
\]
decompactified to describe pointlike branes on a noncompact space in
the presence of a space-filling brane with constant gauge field. We
will now compactify this system (still with a constant gauge field) on
a torus following \cite{wati}. That is, we think of the compactified system as the original branes plus an infinite collection of image branes with the constraint that we keep only configurations invariant under the lattice of translations which
define the torus. 

The full set of branes is described by an infinite matrix which is conveniently broken into finite blocks $X^i_{\vec{n},\vec{m}}$ representing the lowest modes of strings stretched between the $n$th and $m$th set of branes. Similarly, $\Phi$ will be an infinite vector composed of finite vectors $\Phi_{\vec{n}}$ describing the strings stretched between the $n$th set of image branes and the space-filling brane. 

Before imposing the constraint, the full symmetry group of the system is that given at the beginning of section 4,
\beas
{\bf G} \; : \qquad 
&&G_X \; : \qquad X^i \to V X^i V^{-1} \qquad \Phi_k \to \Phi_k V^{-1}
\qquad \cf_a \to \cf_a V^{-1}\\
&&G_A \; : \qquad A_i \to U A_i U^{-1} \qquad \cf_a \to U \cf_a \qquad \Phi_k \to U \Phi_k \\
&&T_X \; : \qquad X^i \to X^i + b^i \qquad \Phi_k \to e^{i k \cdot b} \Phi_k \qquad \cf_a \to \cf_{a-b} \\
&&T_A \; : \qquad A_i \to A_i + l_i \qquad \cf_a \to e^{-i l \cdot a} \cf_a \qquad \Phi_k \to \Phi_{k-l} \; .
\eeas
We would now like to mod out by a subgroup corresponding to the translations that define the torus. More precisely, we wish to restrict to field configurations such that a translation by one of the defining vectors $2 \pi R \vec{k}$ of the torus is equivalent to a gauge transformation which shifts the block indices on $X$ and $\Phi$ by $\vec{k}$. This subgroup is 
\beas
{\bf H} \; : \qquad && X^i \to M_n X^i M_n^{-1} + 2 \pi R n^i \cr
&& A_i \to A_i \cr
&& \Phi_k \to \Phi_k M_n^{-1} e^{i k \cdot (2 \pi R  n)} \cr
&& \cf_a \to \cf_{a - 2\pi R n} M_n^{-1}
\eeas
where for example $(M_n \Phi^\dagger)_m = \Phi^\dagger_{m-n}$. To find the invariant configurations, we note that $H$ maps the subset of variables $\Phi_0$ and $X^i_{0,n}$ one-to one onto the full set of variables $\Phi$ and $X$. Thus, the invariant configurations are specified by choosing $\Phi_0$ and $X^i_{0,n}$ in terms of which the remaining components of $\Phi$ and $X$ determined. To describe the invariant configurations of $\cf$, it is convenient to define
\[
\tilde{\Phi}^\dagger(a,\tilde{y}) = {1 \over (2 \pi \tilde{R})^d} \sum_n (\cf_a)_n e^{-i n \cdot \tilde{y} / \tilde{R}} 
\]
where $\tilde{R} = 2 \pi/R$ is the radius of the dual torus. Then the general invariant configuration of variables is given by
\bea
{\bf \chi_H} \; : \qquad && \{ X| X^i_{m,n} = X^i_{0,n-m} + 2 \pi R m^i \delta_{m,n} \} \cr
&& \{ A \} \cr
&& \{ \Phi_k | (\Phi_k)_n = (\Phi_k)_0 e^{i k \cdot (2 \pi R n)} \} \cr
&& \{ \tilde{\Phi}^\dagger(a,\tilde{y}) | \tilde{\Phi}^\dagger(a+2 \pi
R n,\tilde{y}) = e^{-in \cdot \tilde{y} / \tilde{R}} \tilde{\Phi}^\dagger(a, \tilde{y}) \} \; .
\label{chiH}
\eea
Finally, we may determine the subset $G_H$ of the original symmetry group $G$ which maps this set of invariant configurations into itself. We find that $G_H$ includes all transformations in $G_A$ and $T_X$ as well as transformations in $G_X$ for which $V= e^{i \Lambda}$ with $\Lambda_{m,n} = \Lambda_{0,n-m}$. Finally, although transformations in $T_A$ do not preserve the invariance conditions, the combination of a $T_A$ transformation with parameter $l$ and a $G_X$ transformation with $V_{m,n} = \delta_{m,n} e^{-il \cdot(2 \pi R m)}$ maps $\chi_H$ into itself. The action of these symmetries on the reduced variables is given by
\bea
{\bf G_H} \; : && (G_X)_H \; : \qquad \delta X^i_{0,n} = 2 \pi R i n^i \Lambda_{0,n} + i \Lambda_{0,m} X^i_{0,n-m} -i X^i_{0,m} \Lambda_{0,n-m}\cr 
&&  \qquad \qquad  \qquad \delta (\Phi_k)_0 = -i (\Phi_k)_0 \sum e^{-ik \cdot (2 \pi R n)} \Lambda_{0,n} \cr 
&& \qquad \qquad \qquad \delta \tilde{\Phi}^\dagger(a,\tilde{y}) = -i \tilde{\Phi}^\dagger(a, \tilde{y}) \sum e^{-i n \cdot \tilde{y} / \tilde{R}} \Lambda_{0,n} \cr \cr
&& G_A \; : \qquad A \to U A U^{-1} \qquad \tilde{\Phi}^\dagger(a, \tilde{y}) \to U \tilde{\Phi}^\dagger(a, \tilde{y}) \qquad (\Phi_k)_0 \to U (\Phi_k)_0 \cr \cr
&& T_X \; : \qquad X^i_{0,n} \to X^i_{0,n} + b^i \delta_{0,n} \qquad (\Phi_k)_0 \to e^{i k \cdot b} (\Phi_k)_0 \qquad \tilde{\Phi}^\dagger(a, \tilde{y}) \to \tilde{\Phi}^\dagger(a-b, \tilde{y}) \cr \cr
&& (T_A / G_X)_H \; : \qquad A_i \to A_i + l_i \qquad X^i_{0,n} \to X^i_{0,n} e^{il \cdot(2 \pi R n)} \cr
&& \qquad \qquad \qquad (\Phi_k)_0 \to (\Phi_{k-l})_0 \qquad \tilde{\Phi}^\dagger(a, \tilde{y}) \to e^{-i a \cdot l} \tilde{\Phi}^\dagger(a, \tilde{y}-l)   
\label{comsym}
\eea
The action including the winding modes is now obtained by restricting the original action to the H-invariant configuration space to give a reduced action in terms of $X_{0,n}, (\Phi_k)_0, A, \tilde{\Phi}^\dagger(k,\tilde{y})$. In practice, it is simplest to write the action using the T-dual variables $\tilde{\Phi}^\dagger$ and
\beas
\tilde{X}^i &=& A_i \cr
\tilde{A}_i(\tilde{y}) &=& \sum_n X^i_{0,\vec{n}} e^{-i n \cdot \tilde{y} / \tilde{R}} \cr
\tilde{\cf}^\dagger_a &=& (\Phi_a)_0 \cr
\eeas
These variables describe pointlike branes on the dual torus with winding modes ignored in the presence of space-filling branes with a spatially dependent gauge field. 

From these definitions, we find that 
\[
(h(X^i))_{m,n} \to \int {d \tilde{y} \over (2 \pi \tilde{R})^d} e^{-im \cdot \tilde{y} / \tilde{R}} h(-i \tilde{D}_i(\tilde{y})) e^{in \cdot \tilde{y} / \tilde{R}}
\]
where $\tilde{D}_i(\tilde{y}) = \partial_i + i
\tilde{A}_i(\tilde{y})$. We may now apply this relation to transform the
invariant zero-mode actions to the dual variables. 

For example, bilinear terms (\ref{pact}) involving $\Phi$ in the zero-mode action become 
\beas 
&&\int da \; \tr(\Phi_a {\cal C}(X) \Phi^\dagger_a h(A-a)) \cr
  && \qquad = \int da \; \tr( (\Phi_a)_m ({\cal C}(X))_{m,n} (\Phi^\dagger_a)_n h(A-a)) \cr
&& \qquad \to \int da \; \tr( \tilde{\cf}^\dagger_a e^{ia \cdot(2 \pi R m)} \left( \int {d \tilde{y} \over (2 \pi \tilde{R})^d} e^{-im \cdot \tilde{y} / \tilde{R}} {\cal C}(-i \tilde{D}_i(\tilde{y})) e^{in \cdot \tilde{y} / \tilde{R}} \right) \cr
&& \qquad \hspace{2in} e^{-ia \cdot(2 \pi R n)} \tilde{\cf}_a h(\tilde{X}-a)) \cr
&& \qquad = \int da \; \tr( \tilde{\cf}_a^\dagger (2 \pi \tilde{R})^d \int d \tilde{y} \delta(\tilde{y} - a) {\cal C}(-i \tilde{D}_i (\tilde{y})) \delta(\tilde{y} - a) \tilde{\cf}_a  h(\tilde{X} - a)) \cr
&& \qquad = (2 \pi \tilde{R})^d \delta(0)  \int da \; \tr(\tilde{\cf}_a^\dagger {\cal C}(-i \tilde{D}_i) \tilde{\cf}_a h(\tilde{X} - a))
\eeas  
Here, $\delta(0)$ appears because the original action was invariant under $H$ which maps the reduced variables into themselves. Thus, as a function of the reduced variables $\chi_H$, the original action is redundant and we should divide by the measure of $H$, $\sum_{\vec{n}} = (2 \pi \tilde{R})^d \delta(0)$ to obtain the action for $\chi_H$. Thus, the final transformation from the zero-mode action to the dual action with a spatially dependent gauge field reads  
\be
\label{trans}
\int da \; \tr(\Phi_a {\cal C}(X) \Phi^\dagger_a h(A-a)) \to \int da \; \tr(\tilde{\cf}_a^\dagger {\cal C}(-i \tilde{D}_i) \tilde{\cf}_a h(\tilde{X} - a))
\ee
Note that since $X^i$ appeared in ${\cal C}$ only in commutators, the expression ${\cal C}(-i \tilde{D}_i)$ will be a function of field strengths of $\tilde{A}$ and covariant derivatives of field strengths. 

To complete the description of the action with spatially dependent $A$, we need to provide the relation between $\tilde{\cf}_a$ (the image of $(\Phi_a)^\dagger_0$) and $\tilde{\Phi}= \tilde{\Phi}_0$, the image of $(\cf_0)_0$. This may be obtained from the original relation between $\Phi_k$ and $\cf_a$, 
\beas
\cf_k &=&  e^{-ik \cdot a} \co(A-a,X-k) \Phi_a \cr
%\to  (\cf_a)_0 &=& (\co(A-k, X-a))_{n,0} (\Phi_k)_{n} e^{-ik \cdot a}\cr
%\to  \tilde{\Phi}^\dagger_a &=& \int {d \tilde{y} \over (2 \pi \tilde{R})^d} e^{-i n \cdot \tilde{y} /\tilde{R}} 
%\co(\tilde{X}-k, -i \tilde{D}(\tilde{y}) - a)  e^{ik \cdot n / \tilde{R}} \tilde{\cf}^{\dagger}_k e^{-ik \cdot a}\cr
\Rightarrow \tilde{\Phi}_k^\dagger &=& e^{-i k \cdot a} \co( \tilde{X}-a, -i \tilde{D}(a)-k) \tilde{\cf}^\dagger_a  
\eeas 
Here, the operator $\co$ is to be expanded out with the spatial derivatives in $\tilde{D}_i = \partial_i + i \tilde{A}_i$ acting on gauge fields in other covariant derivatives appearing to the right. For example, using (\ref{leading}), we find that the leading terms in this relation (setting the arbitrary constants in (\ref{leading}) to zero) are
\beas
\Phi^\dagger = \cf^\dagger_a + i (X - a)^i \cf^\dagger_a A_i(a) &-& {1 \over 2} (X-a)^i(X- a)^j \cf^\dagger_a (A_j(a) A_i(a) -i \partial_j A_i(a)) + \cdots
\eeas
Inverting this expression provides a definition of $\cf_a$ in terms of $\Phi$ which generalizes our result for the constant gauge field. In particular, the leading terms obtained in this way for $a=0$ match exactly with the original expression (\ref{covf}) for $\cf$ we obtained in section 2.

We may also choose to start with terms in the zero mode action written in terms of $\cf$. For terms (\ref{fact}) bilinear in $\cf$, this gives 
\beas 
&&\int dk \; \tr(\cf_k h(X-k) \cf^\dagger_k {\cal C}(A) )  \cr
&& \qquad \to \int dk \tr( (\cf_k)_m \left( \int {d \tilde{y} \over (2
\pi \tilde{R})^d} e^{-im \cdot \tilde{y} / \tilde{R}} h(-i
\tilde{D}_i(\tilde{y})-k_i) e^{in \cdot \tilde{y} / \tilde{R}} \right)
(\cf_k)^\dagger_n {\cal C}(A)) \cr
&& \qquad = (2 \pi \tilde{R})^d \int dk \int d \tilde{y} \; \tr(\tilde{\Phi}^\dagger(k,\tilde{y}) \; h(-i \tilde{D}_i(\tilde{y}) - k_i) \; \tilde{\Phi}(k,\tilde{y}) {\cal C} (\tilde{X}))
\eeas  
Here, the covariant derivatives in the expansion of $h$ act both on gauge fields in other covariant derivatives and on $\tilde{\Phi}$. So far, the $k$ integral runs over the whole plane. However, the integrand is actually periodic with respect to k since
\beas
&&\tilde{\Phi}^\dagger(k + n / \tilde{R} ,\tilde{y}) \; h(-i \tilde{D}_i(\tilde{y}) - k_i- n_i /\tilde{R}) \; \tilde{\Phi}(k + n / \tilde{R},\tilde{y})\\
&=& \tilde{\Phi}^\dagger(k ,\tilde{y}) e^{-i n \cdot \tilde{y} / \tilde{R}} \; h(-i \tilde{D}_i(\tilde{y}) - k_i- n_i /\tilde{R}) \; e^{i n \tilde{y} /\tilde{R}}  \tilde{\Phi}(k,\tilde{y})\\
&=& \tilde{\Phi}^\dagger(k ,\tilde{y}) \; h(-i \tilde{D}_i(\tilde{y}) - k_i) \; \tilde{\Phi}(k,\tilde{y})
\eeas
where the second line uses the periodicity properties of $\tilde{\Phi}(k,y)$
obtained in (\ref{chiH}), and the third line uses the fact that $(-i
\partial^{\tilde{y}}_i - n^i / \tilde{R})  e^{i n \cdot \tilde{y} /
\tilde{R}} = e^{in \cdot \tilde{y} / \tilde{R}}
(-i\partial^{\tilde{y}}_i)$.  

The periodicity in $k$ reflects again the fact that we should divide
out by the measure of $H$, so that the final action is expressed in
terms of integrals over the torus and dual torus, 
\be
\label{phikyact}
S_\Phi = (2 \pi \tilde{R})^d \int_{T^d(R)} dk \int_{T^d(\tilde{R})} d
\tilde{y} \; \tr( \tilde{\Phi}^\dagger(k,\tilde{y}) \; h(-i \tilde{D}_i(\tilde{y}) - k_i) \; \tilde{\Phi}(k,\tilde{y}) {\cal C} (\tilde{X}) )
\ee
Again, we need to complete the definition of the action by providing an expression for $\Phi(k,\tilde{y})$ in terms of $\Phi$. We have
\bea
&&\cf^\dagger_k = \co^\dagger(X-k,A-a) \Phi^\dagger_a e^{i k \cdot a} \\
\Rightarrow &&(\cf_k^\dagger)_m = \left( \int {d \hat{y} \over (2 \pi \tilde{R})^d} e^{-im \cdot \hat{y} / \tilde{R}} \co^\dagger(-i \tilde{D}_i(\hat{y}) - k_i, \tilde{X}^i - a^i) e^{i n \cdot \hat{y} / \tilde{R}} \right) e^{-ia \cdot n /\tilde{R}} (\Phi^\dagger_a)_0 e^{i k \cdot a}\cr
\Rightarrow&& {1 \over (2 \pi \tilde{R})^d} \sum_m (\cf_k^\dagger)_m e^{im \cdot \tilde{y} /\tilde{R}} = \int d \hat{y}  \delta(\tilde{y} - \hat{y}) \co^\dagger(-i \tilde{D}_i(\tilde{y}) - k_i, \tilde{X}^i - a^i) \delta(\hat{y} - a) (\Phi^\dagger_a)_0 e^{i k \cdot a}\cr 
\Rightarrow&& \tilde{\Phi}(k,\tilde{y}) = \co^\dagger(-i \tilde{D}_i(\tilde{y}) - k_i, \tilde{X}^i - a^i) \delta(\tilde{y} - a) \tilde{\cf}_a e^{i k \cdot a}
\label{phikydef}
\eea
In this form, $\tilde{\Phi}(k, \tilde{y})$ looks rather singular,
however, in general this expression defines some smooth periodic
function of $\tilde{y}$ (recalling that $(2 \pi \tilde{R})^d
\delta(y-a) = \sum e^{i n \cdot (y-a)/ \tilde{R}}$). The new variable 
$\tilde{\Phi}(k,\tilde{y})$ is related to our original variable
$\tilde{\Phi}_k$ by 
\[
\tilde{\Phi}_k = \int d \tilde{y} \tilde{\Phi}(k,\tilde{y}) \; .
\]

Thus, in the asymmetric situation where we include the spatial dependence of one gauge field but keep only the zero-modes of the dual gauge field, the action may be written in two rather different forms. The first, written in terms of $\cf_a$ involves an integral over the whole covering space with a delta function $\delta(X-a)$ localizing the action to the matrix location $X$. The second, written in terms of the variable $\Phi(k,y)$ involves a double integral over the torus and dual torus. 

To complete this section, we record that the symmetries (\ref{comsym}) written in terms of the dual variables become (dropping tildes)
\beas
{\bf G} \; : \qquad 
&&G_X \; : \qquad X^i \to V X^i V^{-1} \qquad \Phi(k,y) \to \Phi(k,y) V^{-1} \qquad \cf_a \to \cf_a V^{-1} \\
&&G_A \; : \qquad A_i(y) \to U(y) A_i(y) U^{-1}(y) + i \partial_i U(y) U^{-1}(y) \\ && \qquad \qquad \qquad \Phi(k,y)  \to U(y) \Phi(k,y)  \qquad \cf_a \to U(a) \cf_a \\
&&T_X \; : \qquad X^i \to X^i + b^i \qquad \qquad A_i(y) \to A_i (y-b) \\ && \qquad \qquad \qquad \Phi(k,y) \to e^{i k \cdot b} \Phi(k,y-b)  \qquad \cf_a \to \cf_{a-b} \\
&&T_A \; : \qquad A_i \to A_i + l_i \qquad \cf_a \to e^{-i l \cdot a} \cf_a \qquad \Phi(k,y) \to \Phi(k-l,y) \; .
\eeas

\subsection{Full action for pointlike branes with noncompact space-filling brane}

To return to the original scenario discussed in section 2, and in order to obtain the full action on the torus with all winding/momentum modes included, we would now like to decompactify the torus with the spatially dependent gauge field, to yield the action for pointlike branes in the presence of an infinite space-filling brane. 

Using the first form (\ref{trans}) of the action above, the decompactification is trivial. Here, the integration is already over the whole covering space, so the action as written above applies to the non-compact situation as well. Thus, the general invariant bilinear term in $\Phi$ may be written 
\be
\label{fdact}
\int dy \; \tr(\cf^\dagger_y {\cal C}(-i D(y)) \cf_y h(X-a)) 
\ee
where as above,
\[
\Phi^\dagger = \co(X-y, -i D(y)) \cf^\dagger_y
\]

On the other hand, we may start from the second form of the action (\ref{phikyact}), written in terms of $\Phi(k,y)$. In the decompactification limit, the definition (\ref{phikydef}) of $\Phi(k,y)$ is unchanged except that the delta function is no longer periodic. It follows then that the $k$ dependence of $\Phi(k,y)$ becomes simple,
\beas
\Phi(k,y) &=& \co^\dagger(-i D_i(y) - k_i, X^i - a^i) \delta(y - a) \cf_a e^{i k \cdot a} \cr
&=& e^{i k \cdot y} \co^\dagger(-i D_i(y), X^i - a^i) e^{-ik \cdot(y-a)} \delta(y - a) \cf_a \cr
&=& e^{i k \cdot y} \Phi(0,y)
\eeas  
As a result, we find that 
\beas
\Phi^\dagger(k,y) \; h(-i D_i(y) - k_i) \; \Phi(k,y)
&=&\Phi^\dagger(0,y) e^{-i k \cdot y} h(-i D_i(y) - k_i) e^{i k \cdot y} \Phi(k,y) \cr
&=&\Phi^\dagger(0,y) \; h(-i D_i(y)) \; \Phi(0,y)
\eeas
so the integrand of the action (\ref{phikyact}) becomes $k$-independent. The $k$ integral then gives the volume $1/\tilde{R}^d$ of the dual torus, so the decompactified action is
\be
\label{phikydact}
S_\Phi = (2 \pi)^d  \int d y \; \tr( \Phi^\dagger(y) \; h(-i D_i(y)) \; \Phi(y) {\cal C} (X) )
\ee
where we have defined $\Phi(y) \equiv \Phi(0,y)$. 

It is interesting to note that
\be
\label{fourp}
\Phi_k = \int dy \Phi(k,y) = \int e^{i k \cdot y} \Phi(y)
\ee
so that $\Phi(y)$ is simply the Fourier transform of our earlier variable $\Phi_k$, the image of $\Phi$ under a shift $A \to A-k$ in the gauge field.  

As a check, we note that for constant $A$, making the replacement
\[
\Phi(x) = \int {dk \over (2 \pi)^d} e^{-i k \cdot x} \Phi_k \; .
\]
in the action (\ref{phikydact}), gives
\[
S_\Phi = \int dk \; \tr( \Phi_k^\dagger h(A-k) \Phi_k {\cal C}(X))
\]
which is just the zero-mode action we had earlier.

Finally, we note the symmetry transformation rules for the decompactified theory, 
\bea
{\bf G'} \; : \qquad 
&&G_X \; : \qquad X^i \to V X^i V^{-1} \qquad \Phi(y) \to \Phi(y) V^{-1} \qquad \cf_y \to \cf_y V^{-1} \cr
&&G_A \; : \qquad A_i(y) \to U(y) A_i(y) U^{-1}(y) + i \partial_i U(y) U^{-1}(y) \qquad \cf_y \to U(y) \cf_y \cr && \qquad \qquad \qquad \Phi(y) \to U(y) \Phi(y)\cr
&&T_X \; : \qquad X^i \to X^i + b^i \qquad \qquad A_i(y) \to A_i (y-b)\cr && \qquad \qquad \qquad \Phi(y) \to \Phi(y-b)  \qquad \cf_y \to \cf_{y-b} 
\label{gnon}
\eea
Here, we do not need to include separate transformations $T_A$, since these correspond to gauge transformations with $U(x) = e^{ik \cdot x}$. 

\subsection{Full action on a torus}

We now have invariant actions describing pointlike branes on a noncompact space in the presence of a space-filling brane with arbitrary gauge field. By a second application of the quotienting procedure above, we may finally obtain an action for the compactified system with all momentum modes of the gauge field and dual gauge field included.  

This time, the appropriate subgroup of $G'$ for the quotient is  
\beas
{\bf H'} \; : \qquad && X^i \to M_n X^i M_n^{-1} + 2 \pi R n^i \cr
&& A_i(y) \to A_i(y-2 \pi R n) \cr
&& \Phi(y) \to \Phi(y-2 \pi R n) M_n^{-1} \cr
&& \cf_a \to \cf_{a - 2\pi R n} M_n^{-1} \; .
\eeas
To describe the invariant subspace of configurations, we again define
\[
\tilde{\Phi}^\dagger(a,\tilde{y}) = {1 \over (2 \pi \tilde{R})^d} \sum_n (\cf_a)_n e^{-i n \cdot \tilde{y} / \tilde{R}} 
\]
and also 
\[
\tilde{\cf}^\dagger(\tilde{y},a) = e^{ia \cdot \tilde{y}} \sum_n (\Phi(a))_n e^{-i n \cdot \tilde{y} / \tilde{R}} \; . 
\]
Note that $\tilde{\Phi}(a,\tilde{y})$ is a periodic function of $\tilde{y}$ while because of the prefactor in its definition, $\tilde{\cf}(\tilde{y},a)$ satisfies 
\[
\tilde{\cf}^\dagger(\tilde{y} + 2 \pi n \tilde{R},a) = e^{i a \cdot n /R} \tilde{\cf}^\dagger(\tilde{y},a) \; . 
\] 
With these definitions, the invariant subspace of configurations are parameterized by
\bea
{\bf \chi_H} \; : \qquad && \{ X| X^i_{m,n} = X^i_{0,n-m} + 2 \pi R m^i \delta_{m,n} \} \cr
&& \{ A | A(y + 2 \pi R n) = A(y) \} \cr
&& \{ \tilde{\cf}^\dagger(\tilde{y},a) | \tilde{\cf}^\dagger( \tilde{y},a + 2 \pi R n) = \tilde{\cf}^\dagger(\tilde{y},a) \} \cr
&& \{ \tilde{\Phi}^\dagger(a,\tilde{y}) | \tilde{\Phi}^\dagger(a+2 \pi R n,\tilde{y}) = e^{-in \cdot \tilde{y} /\tilde{R}} \tilde{\Phi}^\dagger(a, \tilde{y}) \} \; .
\label{finvar}
\eea
We may now restrict the actions (\ref{fdact}) or (\ref{phikydact}) to this invariant subspace to obtain our final actions for the compactified system. 

From (\ref{fdact}), we find an action
\[
S_\cf = (2 \pi \tilde{R})^d \int dy d \tilde{y} \; \tr( \tilde{\Phi}^\dagger(y, \tilde{y}) h(-i \tilde{D}(\tilde{y}) - y) \tilde{\Phi}(y, \tilde{y}) {\cal C}(-i D(y))) \; .
\]
As discussed above for the action (\ref{phikyact}), the integrand is periodic with respect to $\tilde{y}$, so we restrict the integral over $\tilde{y}$ to the dual torus (thereby dividing out by the measure of $H$).

From (\ref{phikydact}), we find 
\[
S_\Phi = (2 \pi R)^d \int d \tilde{y} dy \; \tr( \tilde{\cf}(\tilde{y}, y) h(-i D(y) - \tilde{y}) \tilde{\cf}^\dagger(\tilde{y},y) {\cal C}(-i \tilde{D}({\tilde{y}}))) \; .
\] 
Again, we find that the integrand is periodic in both variables, so
the integrals in the final expression run over the torus and dual
torus. These actions have identical form up to replacements
\be
\label{td2}
\tilde{\cf}(\tilde{y},y)  \leftrightarrow \tilde{\Phi}^\dagger(y,\tilde{y}) 
\qquad \qquad A(y) \leftrightarrow \tilde{A}(\tilde{y}) 
\ee
as should be expected, since the zero-mode actions from which they
arose were related by T-duality. The replacements (\ref{td2})
generalize our previous T-duality rules to the spatially dependent
degrees of freedom.

To complete the description of the action, we must present a definition of the fields $\tilde{\Phi}(y,\tilde{y})$ and $\tilde{\cf}(\tilde{y},y)$ in terms of an independent degree of freedom. We define first the less redundant variables that appeared in our zero-mode action, 
\beas
\tilde{\cf}^\dagger_a &\equiv& (\Phi_a)_0 \\
\tilde{\Phi}^\dagger_k &\equiv& (\cf_k)_0 \; . 
\eeas
We would then like to express all $\Phi$ and $\cf$ variables in terms of the basic field $\tilde{\Phi} = \tilde{\Phi}_0$.

Firstly, starting from the relation 
\[
\Phi(x) = \co^\dagger(-iD(x),X-a)\delta(x-a) \cf_a
\]
in the noncompact theory, we find that $\tilde{\Phi}(y,\tilde{y})$ and $\tilde{\cf}(\tilde{y},y)$ are related by
\be
\label{rel1}
\tilde{\cf}^\dagger(\tilde{y},x) = (2 \pi \tilde{R})^d e^{i \tilde{y} \cdot x} \int dk \left[ \co^\dagger (-i D(x), -i \tilde{D}(k) -a) \delta(x-a) \delta(k-\tilde{y}) \right] \tilde{\Phi}^\dagger(a,k)
\ee
On the other hand, 
\beas
\Phi^\dagger_k &=& \co(X-a, -i D(a)-k) \cf^\dagger_a e^{-i k \cdot a} \\
&\equiv& \hat{\co}(X-a, A(a)-k) \cf^\dagger_a e^{-i k \cdot a}
\eeas
Here, the first line is defined by expanding the polynomial $\co$ and taking the covariant derivatives (which appear to the right of $\cf^\dagger$) to act on the gauge fields in other covariant derivatives. In the second line   $\hat{\co}$ is simply defined as the result of this procedure and therefore includes explicitly derivatives of its second argument. Formally inverting this expression gives
\[
(\cf_a^\dagger)_n = e^{i k \cdot a} (\hat{\co}^{-1}(X-a,A(a)-k))_{n,m} (\Phi^\dagger_k)_m \; . 
\]
We may now reduce to the invariant subspace of variables (\ref{finvar}) to obtain
\be
\label{rel2}
\tilde{\Phi}(a, \tilde{y}) = e^{i k \cdot a} \hat{\co}^{-1}(-i \tilde{D}(\tilde{y}) -a, A(a) -k) \delta(\tilde{y}-k) \tilde{\cf}_k
\ee
where we have used $(\Phi_k)_n = (\Phi_k)_0 e^{i k \cdot n / \tilde{R}}$. Finally, we may integrate over $\tilde{y}$ to obtain 
\be
\label{rel3}
\tilde{\Phi}_a =  \hat{\co}^{-1}(i \overleftarrow{\partial}_k + \tilde{A}(k) - a, A(a) - k) \tilde{\cf}_k e^{i k \cdot a}\; .
\ee
Here the derivatives $\overleftarrow{\partial_k}$ act to the left on gauge fields appearing in the expansion of $\hat{\co}^{-1}$.

Thus, the relation (\ref{rel3}) defines $\tilde{\cf}_a$ in terms of $\tilde{\Phi} = \tilde{\Phi}_0$ and $\tilde{\Phi}_k$ in terms of $\tilde{\cf}_a$, the relation (\ref{rel2}) to defines $\tilde{\Phi}(a, \tilde{y})$ in terms of $\tilde{\cf}_k$, and the relation (\ref{rel1}) defines $\tilde{\cf}(\tilde{y}, k)$ from $\tilde{\Phi}(a, \tilde{y})$.
From these, or from the definitions above, we have also
\beas
\tilde{\Phi}_k &=& \int d\tilde{y} \tilde{\Phi}(k,\tilde{y}) \cr
\tilde{\cf}_{\tilde{y}} &=& \int da \tilde{\cf}(\tilde{y}, a) \cr
\tilde{\cf}(\tilde{y},a) &=& \sum_l e^{-2 \pi i \tilde{R} l \cdot a} \tilde{\cf}_{\tilde{y} + 2 \pi \tilde{R} l}
\eeas
In the final variables (dropping tildes), the symmetry transformations read 
\bea
{\bf G''} \; : \qquad 
&&G_A \; : \qquad A_i(y) \to U(y) A_i(y) U^{-1}(y) + i \partial_i U(y) U^{-1}(y) \cr
&& \qquad \qquad \qquad \Phi(\tilde{y},y) \to U(y) \Phi(\tilde{y},y) \qquad \cf(y,\tilde{y}) \to U(y) \cf(y,\tilde{y}) \cr
&&G_{\tilde{A}} \; : \qquad \tilde{A}_i(\tilde{y}) \to U(\tilde{y}) A_i(\tilde{y}) U^{-1}(\tilde{y}) + i \partial_i U(\tilde{y}) U^{-1}(\tilde{y}) \cr
&& \qquad \qquad \qquad \cf(y,\tilde{y}) \to  \cf(y,\tilde{y})V^{-1} (\tilde{y}) \qquad \Phi(\tilde{y},y) \to \Phi(\tilde{y},y)V^{-1}(\tilde{y}) \cr
&&T_A \; : \qquad A^i \to A^i + a^i \qquad  \tilde{A}_i(\tilde{y}) \to \tilde{A}_i (\tilde{y}-a)\\ && \qquad \qquad \qquad \Phi(\tilde{y},y) \to \Phi(\tilde{y}-a,y)  \qquad \cf(y,\tilde{y}) \to \cf(y,\tilde{y}-a)e^{-i a \cdot y} \cr
&&T_{\tilde{A}} \; : \qquad \tilde{A}^i \to \tilde{A}^i + \tilde{a}^i \qquad  A_i(y) \to \tilde{A}_i (y-\tilde{a})\cr && \qquad \qquad  \qquad \cf(y,\tilde{y}) \to \cf(y-\tilde{a},\tilde{y}) \qquad \Phi(\tilde{y},y) \to \Phi(\tilde{y},y-\tilde{a}) e^{i \tilde{a} \cdot \tilde{y}}\cr
\label{gpp}
\eea
Here, the $T_{\tilde{A}}$ transformations come from $G_A$ transformations in $G'$ with $U = e^{-il \cdot y}$ combined with $G_X$ transformations with $V_{m,n} = e^{-in \cdot l / \tilde{R}} \delta_{m,n}$.  

\subsection{Summary, Discussion, and Examples}

In this section, we have used a quotienting construction starting from
an invariant zero-mode action to determine the structure of invariant
actions in the case of pointlike branes and noncompact space-filling
branes with spatially dependent gauge field, and in the case of
pointlike and space-filling branes on a torus with all winding and
momentum modes included. We have focused on terms quadratic in the
bifundamentals, however the extension to the more general case is
straightforward. We now summarize and discuss the results.

\subsubsection*{Noncompact case}

We found that a general bilinear term (\ref{pact}) in the zero-mode action leads to a term 
\be
\label{fact2}
S_\cf = \int dy \tr( \cf_y^\dagger {\cal C}(-i D(y))\cf_y h(X-y))
\ee
where $\cf_y$ is related to $\Phi$ by
\[
\Phi^\dagger = \co(X-y, -i D(y)) \cf^\dagger_y \; .
\]
On the other hand, terms (\ref{fact}) written in terms of $\cf_a$ in the zero-mode action lead to terms 
\be
\label{pact2}
S_\Phi = (2 \pi)^d  \int d y \tr( \Phi^\dagger(y) \; h(-i D(y)) \; \Phi(y) {\cal C} (X) )
\ee
where $\Phi(y)$ may be defined in terms of $\cf_a$ by
\[
\Phi(y) = \co^\dagger(-i D(y), X - a) \delta(y - a) \cf_a  \; .
\] 
The Fourier transform (\ref{fourp}) of $\Phi(y)$ reduces to our earlier variable $\Phi_k$ when the gauge field is taken to be constant.

These actions are invariant under all gauge and translation symmetries, with transformation rules given explicitly in (\ref{gnon}). The variables $\Phi(y)$ and $\cf_y$ have identical transformation properties, but a rather different physical interpretation. For abelian $X$, we find 
\[
\cf_y = Pe^{i \int_y^X A}\Phi \qquad \qquad \Phi(y) = \delta(y-X) \Phi
\]
while for general $X$ and $A=0$ we have
\[
\cf_y = \Phi \qquad \qquad \Phi(y) \delta_{sym}(y-X) \Phi 
\]
Thus, $\cf_y$ is something like $\Phi$ parallel transported to the
point $y$ while $\Phi(y)$ is a distribution ``localized'' at the
matrix position of the pointlike branes, something like a projection
or pull-back of $\Phi$ from the noncommutative geometry described by $X$ to the commutative worldvolume of the space-filling branes.

It follows directly from the discussion in section 4.2.4 that we may
obtain a minimal basis of invariant actions either from (\ref{fact2})
or (\ref{pact2}) by taking the matrix function $h$ to be a commutator
expression times a matrix delta function. Now that $A$ has spatial
dependence, this choice also ensures that for diagonal $X$, the actions for
the bifundamentals couple to the gauge field only at points given by
the diagonal elements of $X$. 

A general action may then be written using $\cf$ as a sum of terms of the form
\[
S_\cf = \int dy \; \tr( \cf_y^\dagger \; {\cal C}_1(-i D(y)) \; \cf_y  \; {\cal C}_2(X) \delta(X-y)) \; .
\]
Here ${\cal C}_1(-i D(y))$ will be some general expression built from field strengths and covariant derivatives, and the integral over $y$ is localized to the vicinity of the branes by the matrix delta function. On the other hand, the $\Phi(y)$ basis consists of terms of the form
\be
\label{phinoncat}
S_\Phi = (2 \pi)^d  \int d y \; \tr(\Phi^\dagger(y) \; {\cal C}_2(-i D(y)) \delta(-i D(y)) \; \Phi(y) \; {\cal C}_1 (X)) \; .
\ee
We may process this expression using the relation
\bea
(2 \pi)^d \delta(-iD(y)) \Phi(y) &=& \int dk \; e^{k \cdot D^y} \Phi(y)\cr    
&=& \int dk \; P e^{i \int_y^{y+k} A} \Phi(y+k) \; .
\label{wilson}
\eea
where the Wilson line in the latter expression takes a straight line path. Making this substitution in (\ref{phinoncat}) and making the change of variables $k = x-y$, we find that the action may be written in a bilocal form
\[
S_\Phi = \int dy \int dx \; \tr( \Phi^\dagger(y) \; {\cal C}_2(-i D(y)) P e^{i \int_y^x A}  \; \Phi(x) \; {\cal C}_1 (X)) \; .
\]
Here, the gauge invariant operator ${\cal C}_2(-i D(y))$ is inserted
at one end of a straight Wilson line between two copies of the
variable $\Phi$. This is simply a result of our choice for the minimal
basis; more generally we could have a Wilson line of any shape with
insertions of operators at arbitrary points, but these could be
reexpressed in terms of our minimal basis. Note that in contrast to the $\cf$ action, the $\Phi$ action localizes to the vicinity of the branes because the variable $\Phi(y)$ itself has support here. 

The open Wilson line structures in the $\Phi$ action are reminiscent
of those appearing in the couplings of noncommutative gauge theories
to bulk fields \cite{lm,dt,oo}. This analogy (and the symmetrized property of leading order couplings of D-branes to bulk fields) suggests that a more canonical choice for a minimal basis would be to take
\be
\label{symchoice}
h(X) = sym({\cal C}(X) \delta(X))
\ee
where $sym$ indicates that the individual commutators in ${\cal C}$
should be symmetrized with the powers of $X$ in the integral
definition of the delta function. In the bilocal picture, this means
that the individual field strengths and covariant derivatives of field
strengths will have locations averaged independently over the straight
Wilson line.  
Making this choice of conventions, it follows from the discussion in
section 4.2.4 that we may write the most general invariant bilinear
action consistent with T-duality as a sum of terms 
\bea
S &=&  \int dy dx \; \tr( \Phi^\dagger(y) \; {\cal C}_2(y,x) \; \Phi(x) \; {\cal C}_1 (X)) \cr
&&\qquad + \int dy \; \str(\cf_y^\dagger \; {\cal C}_1(-i D(y)) \;
\cf_y  \; {\cal C}_2(X) \delta(X-y) )
\label{nonctdual} 
\eea
where ${\cal C}_2(y,x)$ is a straight open Wilson line from $y$ to $x$
with the individual commutators in the operator ${\cal C}_2(-i D(y))$
inserted at points which are averaged independently over the
line. Similarly, the symmetrized trace over the $X$'s implements the
choice (\ref{symchoice}) in the second term.

Finally, we note that as a special case, the simple expression
\[
\tr(\cf^\dagger \cf) 
\]
is invariant under all symmetries and consistent with T-duality, since
this decends from the invariant zero-mode action $\tr(\Phi^\dagger
\Phi)$ discussed above. 

%It is interesting to see explicitly the leading terms in the invariant
%completion of a given term in the action for $A=0$, as we did for the
%zero-mode case. For a given leading bilinear term 
%\be
%\label{thisagain}
%\tr(\Phi \Phi^\dagger {\cal C}(X)) \; ,
%\ee
%symmetries and T-duality imply a completion (\ref{nonctdual}) with
%${\cal C}_2 = {\cal C}$ and ${\cal C}_1 = 1$ plus arbitrary collections
%s of higher order terms of the form (\ref{nonctdual}) with both ${\cal
%C}_1$ and ${\cal C}_2$ nonzero. Expanding such an expression in terms
%of $\Phi$, $A$, and $X$, we find that the leading order coupling to
%the gauge field takes the form
%\[
%\int {dk \over (2 \pi)^d} \tr(A_i(-k) J^i(k))
%\]
%where the current $J^i$ is given by
%\[
%J^i(k) = \Phi [X^i, sym( {\cal C} e^{i k \cdot X})] \Phi^\dagger \; 
%\]

\subsubsection*{Compact case}

For pointlike and space-filling branes on a torus, we found that zero-mode actions (\ref{pact}) and (\ref{fact}) lead to terms 
\[
S_\Phi = (2 \pi R)^d \int dy d \tilde{y} \tr(  \Phi^\dagger(\tilde{y}, y) h(-i D(y) - \tilde{y}) \Phi(\tilde{y}, y) {\cal C}(-i \tilde{D}(\tilde{y}))) 
\]
and
\[
S_\cf = (2 \pi \tilde{R})^d \int d \tilde{y} dy  \tr( \cf(y,\tilde{y}) h(-i \tilde{D}(\tilde{y}) - y) \cf^\dagger(y,\tilde{y}) {\cal C}(-i D(y)) )
\] 
for the full action on a torus. In both cases, the action is an integral over both the torus and dual torus. 
Here, $\Phi(y,\tilde{y})$ and $\cf(y, \tilde{y})$ are given in terms of $\Phi$ via (\ref{rel1}, \ref{rel2}, \ref{rel3}). These variables are quasiperiodic, with periodicity properties
\bea
\Phi(\tilde{y} + 2 \pi m \tilde{R}, y + 2 \pi n R) &=& e^{im \cdot y /R} \Phi(\tilde{y}, y) \cr 
\cf( y + 2 \pi n R, \tilde{y} + 2 \pi m \tilde{R}) &=& e^{-in \cdot
\tilde{y} /\tilde{R}} \Phi(\tilde{y}, y) 
\label{per}
\eea
however, as demonstrated above, the integrands of both actions above
are periodic. These
actions are invariant under translations in both tori and the full
spatially dependent gauge groups associated with both the gauge field
and the dual gauge field. The symmetry transformation rules are given
in (\ref{gpp}). 

As for the noncompact case, it follows from our discussion of the
zero-mode action that we may obtain a minimal basis of invariant
actions by choosing $h$ to be a commutator times a matrix delta
function (with the canonical symmetrization (\ref{symchoice}), if desired).
This structure leads again to an interesting bilocal form via a
relation of the form (\ref{wilson}). To obtain the simplest form of the
action, we define\footnote{Note that the periodicity properties of the
hatted and unhatted variables are switched}
\beas
\hat{\Phi}(y, \tilde{y}) &=& e^{-i y \cdot \tilde{y}} \Phi(y,
\tilde{y})\\
\hat{\cf}(\tilde{y},y) &=&  e^{i y \cdot \tilde{y}} \cf(\tilde{y},y)
\eeas
in terms of which the actions become
\beas
S_{\hat{\Phi}} &=& R^d \int dy \; dx \; d \tilde{y} \; \tr(\hat{\Phi}^\dagger(y,\tilde{y}) {\cal
C}_1 (y,x) \hat{\Phi} (x,\tilde{y}) {\cal C}_2(\tilde{y})) \cr
S_{\hat{\cf}} &=& \tilde{R}^d \int d\tilde{y} \; d \tilde{x} \; dy \; \tr(\hat{\cf}(\tilde{y},y) {\cal
C}_1 (\tilde{y},\tilde{x}) \hat{\cf}^\dagger (\tilde{x},y) {\cal C}_2(y)) \; .
\eeas
In this case, ${\cal C}_2$ is a local covariant operator, while 
${\cal C}_1(y,x)$ is a straight Wilson line from
$y$ to $x$ with local covariant expressions inserted at
points which are averaged over the line. Thus, the actions take a form
which is local in one torus and bilocal in the dual torus. 

Finally, by the considerations of section 4.2.4, the most general
invariant action consistent with T-duality may be written as a sum of
terms 
\[
S = S_{\hat{\Phi}}({\cal C}_1, {\cal C}_2) + S_{\hat{\cf}}({\cal C}_1,
{\cal C}_2)  \; .
\]

\section{Adding transverse coordinates}

Given our results for the effective actions describing pointlike and space-filling branes, it is straightforward to understand the effective actions for more general intersecting brane systems by applying T-duality to various directions. Of particular interest is the case where the two sets of branes have some number of common directions. For example, the D1-D5 system has been widely studied in the context of string theory black holes, while the D3-D7 system is important in some flux compactification models of four-dimensional physics. In this section we will treat this case explicitly by considering T-duality in some number of directions transverse to the pointlike and space-filling branes considered so far.\footnote{Of course, the term space-filling is no longer appropriate since these branes were filling only the directions previously under consideration.} For simplicity, we will take the space-filling branes to be noncompact, however the compact case is not essentially different.

At this stage it is important to reintroduce some of the fields we
have ignored so far. Specifically, in addition to the coordinates
$X^i$, the pointlike branes will have additional matrix coordinates
$X^a$ describing their configurations in the transverse directions. In
addition, the branes previously called space-filling will have adjoint
scalar fields $\tilde{X}^a(y)$ describing their transverse
fluctuations. These fields may appear in the invariant actions
(\ref{fdact}) and (\ref{phikydact}) through the arbitrary functions
$h$ and ${\cal C}$. For example, in (\ref{fdact}), the function ${\cal
C}$ may be any covariant expression built from $\{ F_{ij}(a), D_i
F_{jk} (a), \tilde{X}^a, D_i \tilde{X}^a,  \dots\}$ while the function
$h$ may now include arbitrary dependence on both $(X-a)^i$ and $X^a$
(but should still include a factor $\delta(X-a)$ so that the action
localizes for diagonal $X^i$).  

There is now one additional constraint coming from a symmetry we have not yet considered, namely translation invariance in the transverse directions. This requires that the action be invariant under a shift
\[
T_\perp \; :  \qquad X^a \to X^a + b^a \qquad  \tilde{X}^a \to \tilde{X}^a + b^a  \; .
\] 
Invariance under this symmetry translates to a condition 
\be
\label{transinv}
\partial_{X^a} S + \partial_{\tilde{X}^a} S = 0 \; 
\ee
on the action. This will be true if and only if $X^a$ and $\tilde{X}^a$ appear in the action only in commutators $[\tilde{X}^a,\tilde{X}^b], [X^a,X^i], \dots$ or commutator-like terms
\[
\tilde{X}^a(y) \cf_y - \cf_y X^a \qquad \tilde{X}^a(y) \Phi(y) - \Phi(y) X^a   \;.
\]

To add the transverse dimensions, we follow the quotienting procedure above, adding image branes, restricting to the subspace invariant under translations that define a torus, rewriting in terms of the T-dual variables, and decompactifying the dual torus. These steps lead to the usual replacements for which $X^i, A_i, \Phi, \cf$ inherit spatial dependence on the new directions, while commutators involving $X^a$ and $\tilde{X^a}$ give field strengths and covariant derivatives of gauge field $\hat{A}^a$ living on the lower-dimensional branes and new components $A^a$ of the gauge field on the brane which now fills both $i$ and $a$ directions, for example
\beas
{}[X^a, X^b] &\to& -i \hat{F}_{ab}(x)\\
{}[X^a, X^i] &\to& -i \hat{D}_a X^i (x)\\
\tilde{X}^a(y) \cf_y - \cf_y X^a &\to& -i D_a \cf_y(x) \equiv -i(\partial^x_a \cf_y(x) + i A_a(y,x) \cf_y(x) -i \cf_y(x) \hat{A}_a(x))
\eeas

The final result is that the actions describing the brane system with
some common directions are essentially of the form (\ref{fdact}) and
(\ref{phikydact}), but now ${\cal C}$ and $h$ may involve covariant
structures built from $A_a$ and $\hat{A}_a$, and we may have terms
with covariant derivatives acting on $\cf_a$ or $\Phi(y)$.

\subsection{Example: The Dp-D(p+4) System}

As an explicit example of the results implied by this paper, we
consider the supersymmetric Dp-D(p+4) system. We begin with the
Euclidean D(-1)-D3 brane system where the lower dimensional branes are
pointlike. Setting the D3-brane fields to zero, the leading terms in
the effective action are described by the dimensional reduction of
$N=1$ SYM theory in six dimensions with one adjoint hypermultiplet and
$k$ fundamental hypermultiplets, where $k$ is the number of
D3-branes. Using the conventions of \cite{odb} the action is given by 
\beas
S &=& \tr \left({1 \over 4} [X^a, X^b][X^a,X^b] + {1 \over 2} [X^a,
\bar{X}^{\rho \dot{\rho}}][X^a, X_{\rho \dot{\rho}}] - {1 \over 4}
[\bar{X}^{\a \da}, X_{\b \da}][\bar{X}^{\b \db}, X_{\a \db}] \right)\\
& & + \tr \left({1 \over 2} \bl^\r \gamma^a [X^a , \lambda_\r] + {1 \over
2} \bt^\da \gamma^a [X^a, \theta_\da] - \sqrt{2} i \e^{\a \b} \bt^\da
[X_{\b \da}, \lambda_\a] \right) \\
&& + {\rm tr} \left(\bP^\a [\bX^{\b 
\da}, X_{\a \da}] \P_\b - \bP^{\r} X^a X^a \P_\r \right)\\
&& + {\rm tr} \left(  \bc \gamma^a X^a \chi + \sqrt{2} i \e^{\a \b} \bc 
\lambda_\a \P_\b - \sqrt{2} i \e_{\a \b} \bP^\a \bl^\b \chi \right)\\
&& + {\rm tr} \left(  {1 \over 2}\bP^\a \P_\b \bP^\b \P_\a - 
\bP^\a \P_\a \bP^\b \P_\b \right)
\eeas
Here, undotted and dotted Greek indices are fundamental indices of
$SU_R(2)$ and $SU_L(2)$ which make up the $SO(4)$ rotations in the
directions of the D3-brane, so that the scalars $X^i$ appear here as
$X_{\a \da}$.

The bifundamental terms appear in the last three lines of this action.
In this system, we have two sets of bifundamental fields, scalars
$\Phi_\rho$ transforming in the fundamental of $SU_R(2)$ and fermions
$\chi$. In writing an invariant completion, we will thus require
separate fields analogous to $\cf_a$ or $\Phi(a)$ for each of these,
but everything we have said goes through unchanged in the case with
more than one type of bifundamental field, so it is straightforward to
apply our results and write down the general invariant completion of
this action consistent with T-duality. 

While there are arbitrary parameters in this completion, certain
leading terms are predicted uniquely, and we now give a few examples
of these. For simplicity, we focus on terms involving only the
zero-modes of the fields on the higher dimension branes. 

As a first example, consider the term
\[
\bar{\chi} \gamma^a X^a \chi
\]
From equation (\ref{biglong}), the invariant completion of this term 
consistent with T-duality contains leading order terms
\[
\bar{\chi} \gamma^a (X^a \chi - \chi \tilde{X}^a) -i \bar{\chi}
\gamma^a [X^a, X^i] \chi A_i + \dots \; .
\]
These terms obey the condition (\ref{transinv}) demanded by
translation invariance in the transverse directions, so we may apply
T-duality to obtain the corresponding terms in the general Dp-D(p+4)
action. For the D5-D9 case, we find
\[
\int d^6 x \; \bar{\chi} \gamma^a D_a \chi - \bar{\chi} \gamma^a D_a
X^i \chi A_i + \dots
\]
where $D_a \chi = \partial_a \chi + i \hat{A}_a \chi - i \chi
A_a$. An interesting check is provided by taking abelian $X$, for
which this reduces to 
\[
\int d^6 x \; \bar{\chi} \gamma^a (\partial_a \chi + i \hat{A}_a \chi - i
\chi A_a - \partial_a X^i \chi A_i)
\]
Here, we see that the entire expression in brackets is just
the covariant derivative with respect to both the gauge field on the
D5-branes and the pull-back of the full D9-brane gauge field, as we would
expect. Thus the more general result (including higher order terms we
have not written) should provide a generalization of this
pull-back to the case of nonabelian branes. A similar structure arises
in the invariant completion of the $\bP^\r X^aX^a \P_\r$ term.

In a similar way, we may determine the leading couplings to higher
dimensional brane fields following by
symmetries from all of the other bifundamental terms in the action
above. Still restricting to zero-modes of the higher brane fields, we
find that the linear couplings to $A_i$ and $\tilde{X}^a$ in the
D(-1)-D3 action take the form 
\[
S = \tr(A_i J^i) + \tr(\tilde{X}^a J_a) 
\]
where 
\beas 
J^i &=& i \bP^\a [X^i,[\bX^{\b \da}, X_{\a \da}]] \P_\b -i\bP^\r [X^i,
X^a X^a] \P_\r \cr
&& + i \bP^\a[X^i, \P_\b \bP^\b] \P_\a - 2i \bP^\a [X^i, \P_\a \bP^\b]
\P_\b \cr
&&-i \bc \gamma^a [X^a, X_i] \chi - \sqrt{2} \e^{\a \b} \bc [X^i, \lambda_\a]
\P_\b + \sqrt{2} \e_{\a \b} \bP^\a[X^i, \bl^\b] \chi \cr
J_a &=& 2 \bP^\r X^a \P_\r - \bc \gamma^a \chi
\eeas
The analogous currents for the other Dp-D(p+4) systems may be found by 
T-dualizing these expressions.

\section{Discussion}

In this paper, we have carefully studied one example of how to
implement local symmetries acting on fields associated with D-branes
described by matrix coordinates. We have found that the action of
these symmetries and the form of the resulting invariant actions are quite
complicated when written in terms of the original variables, but that
actions and transformation rules become simple upon defining new
covariant variables. These covariant variables promote the original
bifundamental fields associated with the lower dimensional branes to
(highly redundant) fields defined over the whole worldvolume of the
higher-dimensional branes, and the resulting actions take the form of
an integral over this worldvolume. This is really what we should have expected, since 
general matrix configurations correspond to delocalized D-branes that should not be expected to couple to the gauge field only at a finite collection of points.

The full requirements of symmetry and T-duality on the brane systems
we have considered are quite constraining. We have seen that a given
structure invariant at leading order must be completed by an infinite
series of higher order terms in the $\alpha'$ expansion. While there
is significant freedom in this completion corresponding to adding
arbitrary higher-order invariant structures, we have seen that some of
the leading corrections to the known low energy actions, specifically certain
linear couplings to the gauge field, are determined uniquely. 

One significant open problem is to find
a full solution for the operator $\co$ which defines the relationship
between $\Phi$ and $\cf$ (and thereby enacts a T-duality
transformation) in the zero-mode action and prove that the solution is
unique up to the field redefinitions described in section 4.1.

\subsubsection*{Application to matrix models of M-theory with M5-branes}

An interesting application of our results is to the Berkooz-Douglas
matrix model describing M-theory in the presence of longitudinal
M5-branes. This theory is obtained as the low-energy, weak coupling
limit of the D0-D4 system in which only the massless degrees of
freedom arising from 0-0 and 0-4 strings remain. The usual theory
describes the M5-branes with trivial worldvolume fields, while an
extension to include weak background bulk fields of 11-dimensional
supergravity was given in \cite{odb}. The results of this paper may be
used to obtain a similar extension to include a weak background of
worldvolume M5-brane fields. Specifically, one may use the results of
section 6.1 for the leading order couplings of D0-branes to D4-brane
worldvolume fields together with the relationship between D4-brane and
M5-brane worldvolume fields to obtain leading order couplings. 

\subsubsection*{Extension to D-branes in curved spaces}

Finally, it is interesting to compare our results to those of de Boer and Schalm for D0-branes in a curved space. We have seen that extending the local gauge symmetry to the bifundamental fields requires a complicated gauge transformation for $\Phi$ depending explicitly on the value of the gauge field, and that while invariant actions exist, they are complicated infinite power series when written in terms of $\Phi$. The story in \cite{ds} is very similar. The authors found that extending the general coordinate transformations to the matrix $X$ requires a complicated transformation law depending explicitly on the value of the metric. They were able to explicitly construct invariant actions as a power series in $X$, but these looked quite complicated. 

In our case, the story became much nicer with the discovery of
covariant generalizations $\cf_a$ and $\Phi(a)$ of $\Phi$, behaving
respectively like a parallel transported version of $\Phi$ over the
whole space, and a projection of $\Phi$ with support in the vicinity
of the branes. Given these covariant objects, it became possible to
write simple expressions for fully invariant actions.  

It is certainly conceivable that a similar picture may apply to the
case of D-branes in a metric. This would mean that the transverse
D-brane coordinate matrices $X^i$ could be generalized to fields
$X^i(y)$ defined over the whole transverse space, where $X^i(y)$ would
be determined in terms of $X^i$ and the metric $h_{ij}$. These fields
could either have properties similar to $\cf_a$ (i.e. a parallel
transported version of $X$ reducing to a constant field $X^i(y) = X^i$
in flat space) or similar to $\Phi(a)$ (having support in the vicinity
of the branes) or there could exists objects of both types. 
In either case, the resulting fields should transform
in some simple covariant way under coordinate transformations,
allowing simple expressions for invariant actions (involving an
integral over space). Whether such a picture actually exists is
currently under investigation. 

\vskip 1cm

\noindent
{\large \bf Acknowledgments}\\
\\
I would like to thank Michael Douglas, Clifford Johnson, Rob Myers, and Wati Taylor for helpful discussions. The work of M.V.R is supported in part by NSF grant PHY-9870115 and in part by funds from the Stanford Institute for Theoretical Physics. The work of M.V.R. is also supported by NSERC grant  and by the Canada Research Chairs programme. 

\vskip 1cm

\appendix

\section{Solving the constraints}

In this appendix, we provide an algorithm to determine the general solution for the operator $\co$ which determines $\cf$ in terms of $\Phi$ in the zero-mode sector. 

We defined $\co$ by the expansion
\beas
\cf = {\cal O}(A, X) \Phi &=& \sum_n \sum_{\sigma_n} c_{\sigma_n} A_{i_{\sigma(1)}} \cdots A_{i_{\sigma(n)}} \Phi X^{i_n} \cdots X^{i_1} \cr
&=& \Phi + c_1 A_{i_1} \Phi X^{i_1} + c_{12} A_{i_1} A_{i_2} \Phi X^{i_2} X^{i_1} + c_{21} A_{i_2} A_{i_1} \Phi X^{i_2} X^{i_1} + \dots 
\eeas
and we would now like to determine the coefficients $c_{\sigma_n}$. 

\subsubsection*{T-duality constraint}

The first constraint is that $\co$ should be unitary, 
\be
\label{U2}
\co^{-1} = \co^\dagger
\ee
This may be rewritten directly in terms of the coefficients $c_{\sigma_n}$ as
\be
\label{c1}
c^\dagger_{\sigma} = c^{-1}_\sigma
\ee
where 
\be
\label{def1}
c^\dagger_{\sigma} = c^*_{\sigma^{-1}}
\ee
and
\be
\label{def2}
c^{-1}_{i_1 \cdots i_n} = \sum {}' (-1)^{k+1} c_{i_1 \cdots i_{l_1}} c_{(i_{l_1+1} - l_1) \cdots (i_{l_2} - l_1)} \cdots c_{(i_{l_{k-1} + 1} - l_{k-1}) \cdots (i_n - l_{k-1})} \; .
\ee
Here, the sum on the right includes only terms for which all indices are positive integers. In this sum, we have one term for each way of inserting partitions into the ordered set $(i_1, \cdots ,i_n)$ such that the elements in each group are all greater than the elements in the previous groups. With these relations, the unitarity constraint translates to a set of linear equations for the coefficients at a given order with constants given by sums of products of the lower order terms. For example, at third order, we find
\[
\ba{lllllll}
c^\dagger_{123} &= &c^{-1}_{123}  &\qquad \Rightarrow \qquad& c^*_{123} &=& -c_{123} + c_{12} c_1 + c_{1} c_{12} - c_1 c_1 c_1 \cr
c^\dagger_{132} &= &c^{-1}_{132}  &\qquad \Rightarrow \qquad& c^*_{132} &=& -c_{132} + c_{1} c_{21}\cr
c^\dagger_{213} &= &c^{-1}_{213}  &\qquad \Rightarrow \qquad& c^*_{213} &=& -c_{213} + c_{21} c_3\cr
c^\dagger_{231} &= &c^{-1}_{231}  &\qquad \Rightarrow \qquad& c^*_{312} &=& -c_{231}\cr
c^\dagger_{312} &= &c^{-1}_{312}  &\qquad \Rightarrow \qquad& c^*_{231} &=& -c_{312}\cr
c^\dagger_{321} &= &c^{-1}_{321}  &\qquad \Rightarrow \qquad& c^*_{321} &=& -c_{321} 
\ea
\]

\subsubsection*{Symmetry constraints}

Additional relations come from the constraint
\[
\partial_{A_j} ( \ca_i ) = \delta_{ij} \qquad \qquad 
\ca_i \equiv -i \partial_{X^i} \co \co^{-1}
\]
which came from demanding consistency of the symmetry transformation rules.

To solve this constraint, we first rewrite the definition of $\ca$ as  
\[
(\partial_{X^i} - i \ca_i) \co = 0 \; .
\]
Using a superscript to denote the order (in the $X$ expansion) of $\co$ and $\ca$, this may be rearranged to give
\be
\label{obyo}
\ca_i^n = -i \partial_{X^i} \co^n - \sum_{l=1}^{n-1} \ca_i^l \co^{n-l}
\ee
In general, we then have 
\[
\ca^n_i \Phi = \sum_{\sigma_n} b^k_{\sigma_n} A_{i_{\sigma(1)}} \cdots A_{i_{\sigma(k-1)}} A_i A_{i_{\sigma(k)}} \cdots A_{i_{\sigma(n)}} \Phi X^{i_n} \cdots X^{i_1}
\]
where the coefficients $b^k_{\sigma_{n-1}}$ are determined from (\ref{obyo}) in terms of the $c_{\sigma_{k \le n}}$ by the recursion relations
\be
\label{def3}
b^k_{i_1 \cdots i_{n-1}} = -i \sum_{l=1}^n c_{\bar{i}_1 \cdots \bar{i}_k l \bar{i}_{k+1} \cdots \bar{i}_{n-1}} - \sum_{l=1}^{n-1} b^k_{i_1 \cdots i_{l-1}} c_{(i_l-l+1) \cdots (i_{n-1}-l+1)}   
\ee
In the first sum here, $\bar{i}$ is $i$ for $i<l$ or $i+1$ otherwise.

Given this definition, the constraints on the coefficients $c_\sigma$
arising from the relation $\partial_{A_i} \ca_j^{n > 0} = 0$ are 
\be
\label{c2}
0 = \sum_k b^k_{\sigma} 
\ee
for any $\sigma$ and
\be
\label{c3}
0 = b^k_{l \bar{\sigma}(1) \cdots \bar{\sigma}(n)} + \cdots + b^k_{\bar{\sigma}(1) \cdots \bar{\sigma}(k-1) l  \bar{\sigma}(k) \cdots \bar{\sigma}(n)} + b^{k-1}_{\bar{\sigma}(1) \cdots \bar{\sigma}(k-1) l  \bar{\sigma}(k) \cdots \bar{\sigma}(n)} + \cdots + b^{k-1}_{ \bar{\sigma}(1) \cdots \bar{\sigma}(n) l}
\ee
for any $1 \le k,l \le n+1$ and $\sigma$. Again, $\bar{i}$ denotes $i$ for $i<l$ and $i+1$ otherwise. These two equations just express the fact that $\ca_i^n$ should involve $A_i$ only in commutator expressions. Like the T-duality constraint, this symmetry constraint gives a set of linear equations for the coefficients at a given order, with the constant terms given by sums of products of the lower order coefficients.

\subsubsection*{Summary}

To summarize, given the definitions (\ref{def1}, \ref{def2}, \ref{def3}), the T-duality and symmetry constraints demand that the coefficients $c_\sigma$ satisfy the conditions (\ref{c1},\ref{c2},\ref{c3}). Assuming these have been solved to some order $n-1$, the $n$th order constraints give a set of linear equations for the coefficients $c_{\sigma_n}$. In general, many of these equations are redundant, and by implementing the algorithm above using Maple, we have found (to fifth order in $X$) that there are solutions with a certain number of arbitrary coefficients.

\end{document}